\documentclass[manuscript]{aastex62}

\usepackage{hyperref}
\usepackage{multirow}
\graphicspath{{./}{figures/}}

\accepted{2020 May 11}
\submitjournal{ApJ}

\shorttitle{AGILE Monte Carlo simulation}
\shortauthors{Fioretti et al.}

\begin{document}

\title{AGILESim: Monte Carlo simulation of the AGILE gamma-ray telescope}

\correspondingauthor{Valentina Fioretti}
\email{valentina.fioretti@inaf.it}

\author[0000-0002-6082-5384]{V. Fioretti}
\affil{INAF OAS Bologna, Via P. Gobetti 101, 40129 Bologna, Italy}

\author{A. Bulgarelli}
\affil{INAF OAS Bologna, Via P. Gobetti 101, 40129 Bologna, Italy}

\author{M. Tavani}
\affil{INAF IAPS, Via del Fosso del Cavaliere 100, 00133 Roma, Italy}

\author{S. Sabatini}
\affil{INAF IAPS, Via del Fosso del Cavaliere 100, 00133 Roma, Italy}

\author{A. Aboudan}
\affil{CISAS, Via Venezia 15, 35131 Padova, Italy}

\author{A. Argan}
\affil{INAF IAPS, Via del Fosso del Cavaliere 100, 00133 Roma, Italy}

\author{P. W. Cattaneo}
\affil{INFN Pavia, Via A. Bassi 6, 27100 Pavia, Italy}

\author{A. W. Chen}
\affil{School of Physics, Wits University, Johannesburg, South Africa}

\author{I. Donnarumma}
\affil{INAF IAPS, Via del Fosso del Cavaliere 100, 00133 Roma, Italy}
\affil{Agenzia Spaziale Italiana (ASI), Via del Politecnico snc, 00133 Roma, Italy}

\author{F. Longo}
\affil{Department of Physics, University of Trieste, Via Valerio 2, 34127 Trieste, Italy}
\affil{INFN Trieste, Via Valerio 2, 34127 Trieste, Italy}

\author{M. Galli}
\affil{ENEA Bologna, Via Martiri di Monte Sole 4, 40129 Bologna, Italy}

\author{A. Giuliani}
\affil{INAF IASF Milano, Via E. Bassini 15, 20133 Milano, Italy}

\author{M. Marisaldi}

\affil{Birkeland Centre for Space Science, Department of Physics and Technology, University of Bergen, Norway}
\affil{INAF OAS Bologna, Via P. Gobetti 101, 40129 Bologna, Italy}

\author{N. Parmiggiani}
\affil{INAF OAS Bologna, Via P. Gobetti 101, 40129 Bologna, Italy}
\affil{Università degli Studi di Modena e Reggio Emilia, DIEF - Via Pietro Vivarelli 10, 41125 Modena, Italy}

\author{A. Rappoldi}
\affil{INFN Pavia, Via A. Bassi 6, 27100 Pavia, Italy}



\begin{abstract}

The accuracy of Monte Carlo simulations in reproducing the scientific performance of space telescopes (e.g. angular resolution) is mandatory for a correct design of the mission. A brand-new Monte Carlo simulator of the Astrorivelatore Gamma ad Immagini LEggero (AGILE)/Gamma-Ray Imaging Detector (GRID) space telescope, \textit{AGILESim}, is built using the customizable Bologna Geant4 Multi-Mission Simulator (\textit{BoGEMMS}) architecture and the latest Geant4 library to reproduce the instrument performance of the AGILE/GRID instrument.  The Monte Carlo simulation output is digitized in the \textit{BoGEMMS} postprocessing pipeline, according to the instrument electronic read-out logic, then converted into the onboard data handling format, and finally analyzed by the standard mission on-ground reconstruction pipeline, including the Kalman filter, as a real observation in space.  In this paper we focus on the scientific validation of \textit{AGILESim}, performed by reproducing (i) the conversion efficiency of the tracker planes, (ii) the tracker charge readout distribution measured by the on-ground assembly, integration, and verification activity, and (iii) the point-spread function of in-flight observations of the Vela pulsar in the 100 MeV -- 1 GeV energy range. We measure an in-flight angular resolution (FWHM) for Vela-like point sources of $2.0^{+0.2}_{-0.3}$ and $0.8^{+0.1}_{-0.1}$ degrees in the 100 - 300 and 300 -- 1000 MeV energy bands, respectively.
The successful cross-comparison of the simulation results with the AGILE on-ground and in-space performance validates the \textit{BoGEMMS} framework for its application to future gamma-ray trackers (e.g. e-ASTROGAM and AMEGO).

\end{abstract}


\section{Introduction}

The observation in space of the gamma-ray sky is mainly carried out today in the pair production regime, from tens of MeV to tens of GeV, by AGILE (launched by ASI in 2007, \citet{2009A&A...502..995T}) and Fermi (launched by NASA in 2008, \citet{2009ApJ...697.1071A}). Both telescopes share the common design of the current generation pair conversion telescopes: a tungsten-silicon tracker, a calorimeter based on crystal scintillators at the bottom, and a plastic scintillator as anticoincidence system (ACS) surrounding the two instruments. The gamma-rays interact with the tracker layers, producing an electron-positron pair that is tracked along its path by the Silicon microstrip detectors composing the tracker trays. The direction of the pair, together with the energy collected by the calorimeter, allows the reconstruction of the energy and direction of the primary gamma-ray. The ACS vetoes the tracks produced by charged particles, which outnumber the gamma-rays by a factor of $10{^4}$ and more. 
\\
The energy range of pair conversion telescopes is limited from below by the increasing multiple Moli\`{e}re scattering angles of electrons and positrons in the tracker layers (mainly Tungsten).
The energies below 30-50 MeV, where the Compton scattering is the dominant interaction process, are not covered by current tracking telescopes.
Although the coded IBIS imager on board the ESA INTEGRAL \citep{2003A&A...411L...1W} mission, launched in 2002, reaches an energy upper limit of 10 MeV, its continuum sensitivity is a factor of $10^{3}$ worse than AGILE and Fermi above 1 MeV because of the transparency of the Tungsten coded mask and the low attenuation of the CsI scintillator \citep{2003A&A...411L.131U}. The best, though modest, sensitivity of $6\times10^{-5}$ ph. cm$^{-2}$ s$^{-1}$ in the 1 - 30 MeV band \citep{2004NewAR..48..193S} has been achieved in the past by COMPTEL \citep{1993ApJS...86..657S}, the first and only space Compton telescope that was flown onboard the NASA Compton Gamma-ray Observatory (CGRO) from 1991 to 2000.
\\
After thousands of gamma-ray sources discovered by the AGILE and Fermi telescopes \citep[see][]{2015ApJ...810...14A}, with many of them still to be identified, large efforts are currently undertaken to fill the observational gap in the MeV domain by conceiving new technological solutions, and at the same time, to increment the performance of present converting trackers in view of the future Cherenkov Telescope Array (CTA) observatory \citep{2011ExA....32..193A}. 
\\
A common requirement for the next generation gamma-ray missions is the development of scientifically validated Monte Carlo simulations to explore and test new designs and analysis algorithms. Although each launch in space is anticipated by extended simulations of the spacecraft and instrument interaction with the radiation environment and the science targets, very rarely the simulation framework is updated \textit{a posteriori} according to the real in-flight data sets. We present here the design (Sec. \ref{sec:design}) and the full validation chain (Sec. \ref{sec:validation}) of the \textit{AGILESim} framework, the Geant4 Monte Carlo simulator of the AGILE mission based on the \textit{BoGEMMS} (Bologna Geant4 Multi-Mission Simulator) architecture \citep{2012SPIE.8453E..35B}.
Being able to reproduce the AGILE/GRID in-flight angular resolution serves many purposes: (i) proving the reliability of the \textit{BoGEMMS} gamma-ray simulation branch \citep{2014SPIE.9144E..3NF} as an accurate tool to evaluate the scientific requirements (e.g. angular resolution, sensitivity) of future gamma-ray missions, (ii) obtaining the ultimate calibration of the in-flight AGILE operations and confirming its excellent angular resolution, (iii) validating the Geant4 toolkit as the simulation code of reference for gamma-ray space applications.

\section{The AGILE mission}
The AGILE (Astrorivelatore Gamma ad Immagini LEggero, \citet{2019RLSFN.tmp...38T}) gamma-ray satellite, with more than ten years of activity, operates below the Van Allen belts in an equatorial low Earth orbit at an altitude, at launch, of $\sim550$ km, where the geomagnetic cut-off ensures a low level of particle background. The telescope operated in pointing mode during the period April 2007 -- October 2009 and until now in spinning mode around its solar panel axis. 
\\
The AGILE scientific payload comprises the Gamma-Ray Imaging Detector (GRID) for observations in the 30 MeV - 30 GeV energy range and a coded X-ray detector (Super-AGILE, \citet{2007NIMPA.581..728F}), currently working in rate-meter mode in the 20 - 60 keV energy range because of temporary telemetry limitations \citep{2016ApJ...825L...4T}. 
\\
The AGILE/GRID instrument is an electron tracking telescope consisting of a Silicon Tracker (ST, \citet{2010NIMPA.614..213B}, a Cesium Iodide (CsI(Tl)) Mini-Calorimeter (MCAL) sensitive in the 300 keV - 100 MeV energy range \citep{2009NIMPA.598..470L}, and a surrounding Bicron BC-404 plastic scintillator acting as anticoincidence system (ACS, \citet{2006NIMPA.556..228P}). The ST consists of 13 mechanical structures (trays) vertically positioned on top of each other. 
Each tray is composed of a load-bearing structure made of Aluminum honeycomb, two intermediate Carbon fiber layers and two Silicon strip layers, except for the two outermost trays (first and last) which only have one. In such a way, 12 XY planes are available, each composed of two Silicon layers placed directly one facing the other, with the respective strips oriented along the X and Y directions of the telescope. The first 10 trays from the top (or sky side) feature a Tungsten foil for the photon conversion into an electron-positron pair, the threshold for the conversion being two times the electron rest mass. 
Each Si layer is equipped with interleaved \textit{floating} and \textit{readout} strips (see Sec. \ref{sec:pp}).
The sequence of Si planes tracks the trajectories of the electron/positron pair, which then interacts with the calorimeter and exits the telescope. 
The gamma-ray direction is reconstructed by means of the track directions measured in the tracker. The energy is instead measured from the multiple scattering of the pair in the Si planes. The MCAL is too small to collect the full energy of the pair, but its events are used by the on-board filter to better discriminate true gamma-ray events. 
The GRID energy range is limited at low energies by the multiple scattering of the tracks in the planes, that reduces the tracking efficiency, and the Tungsten self-absorption. The range upper limit is instead caused by the lower reconstruction efficiency to gamma-rays because of the small aperture angle of the pair, causing a significant decrease in the effective area.
The MCAL is simulated here as instrument complementary to the tracking telescope, but it can operate stand-alone in burst mode for the observation of soft gamma-ray events as, e.g., gamma-ray bursts \citep{2013A&A...553A..33G} and terrestrial gamma-ray flashes \citep{2014JGRA..119.1337M}.
\\
The GRID design, the analog trigger logic, and the dedicated data handling and acquisition system allow to achieve \citep{2011NIMPA.630....7T}, with a very light ($\sim100$ kg) instrument, (i) an excellent angular resolution, comparable to Fermi/LAT (Large Area Detector) in the overlapping energy range \citep{2015ApJ...809...60S}; (ii) a very large field of view ($\sim2.5$ sr), 5 times larger than EGRET; (iii) a good sensitivity to transient emission of $(1-2)\times10^{-8}$ erg cm$^{-2}$ s$^{-1}$ \citep{2016ApJ...825L...4T} at energies above 30 MeV for 100 s integration time (spinning mode).
\\
%
%
\begin{figure*}
   \centering
    \includegraphics[trim=0cm 5.cm 0cm 5.cm, clip=true,width=0.99\textwidth]{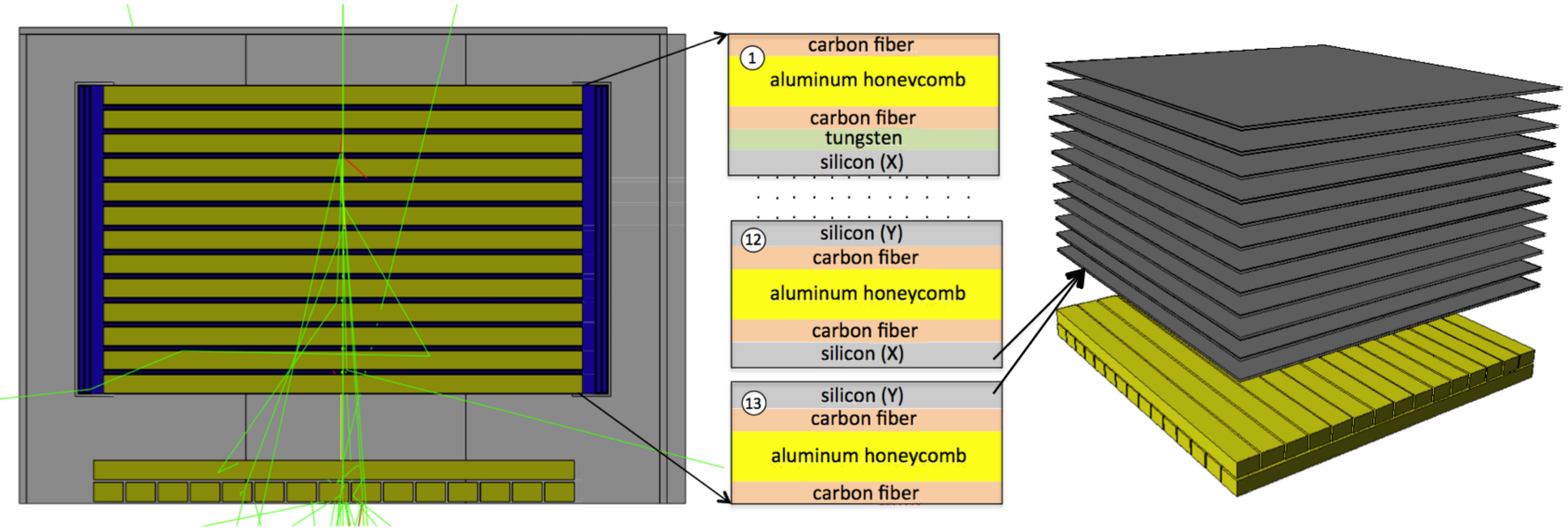}  
    \includegraphics[width=0.33\textwidth]{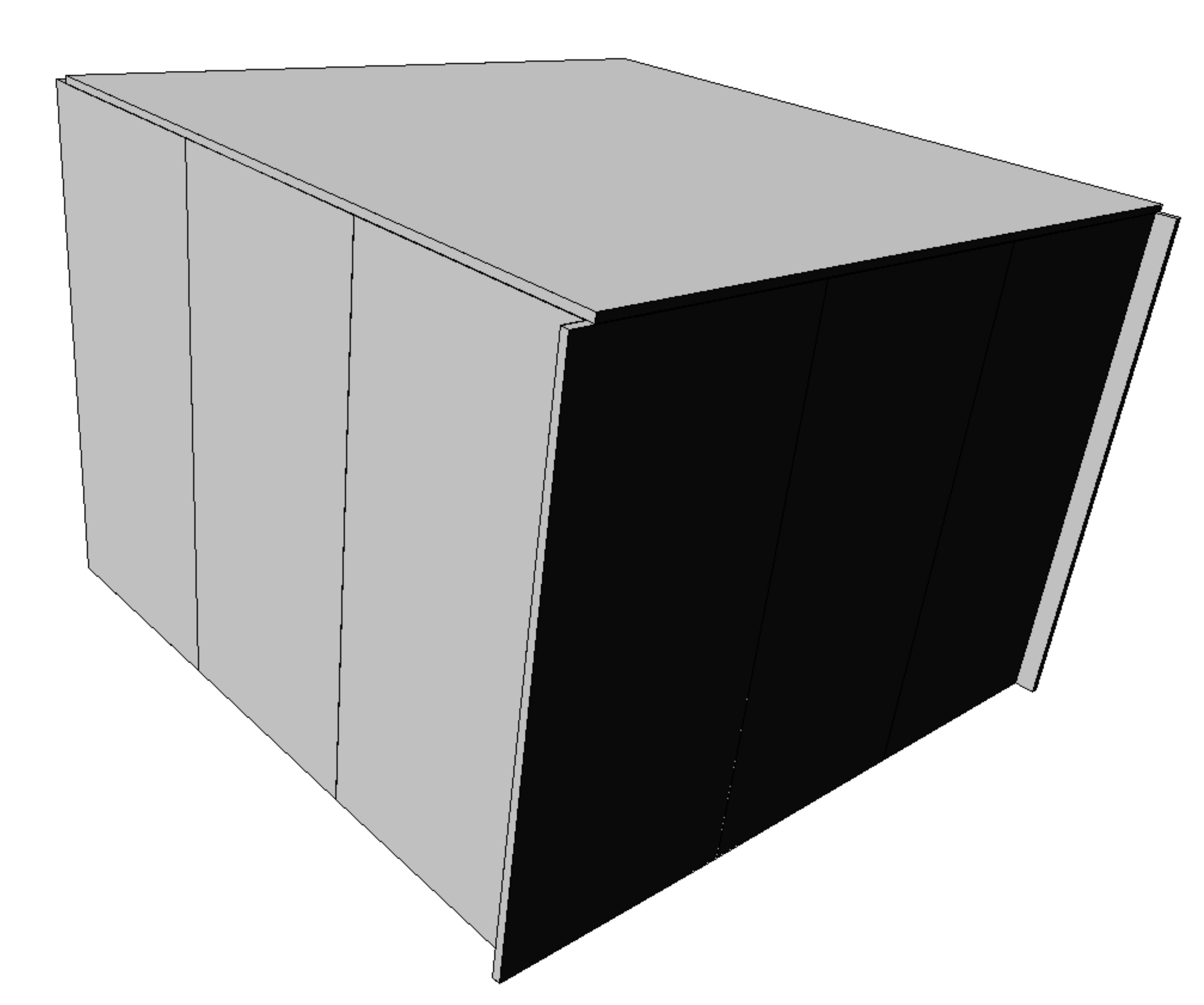}
   \includegraphics[width=0.33\textwidth]{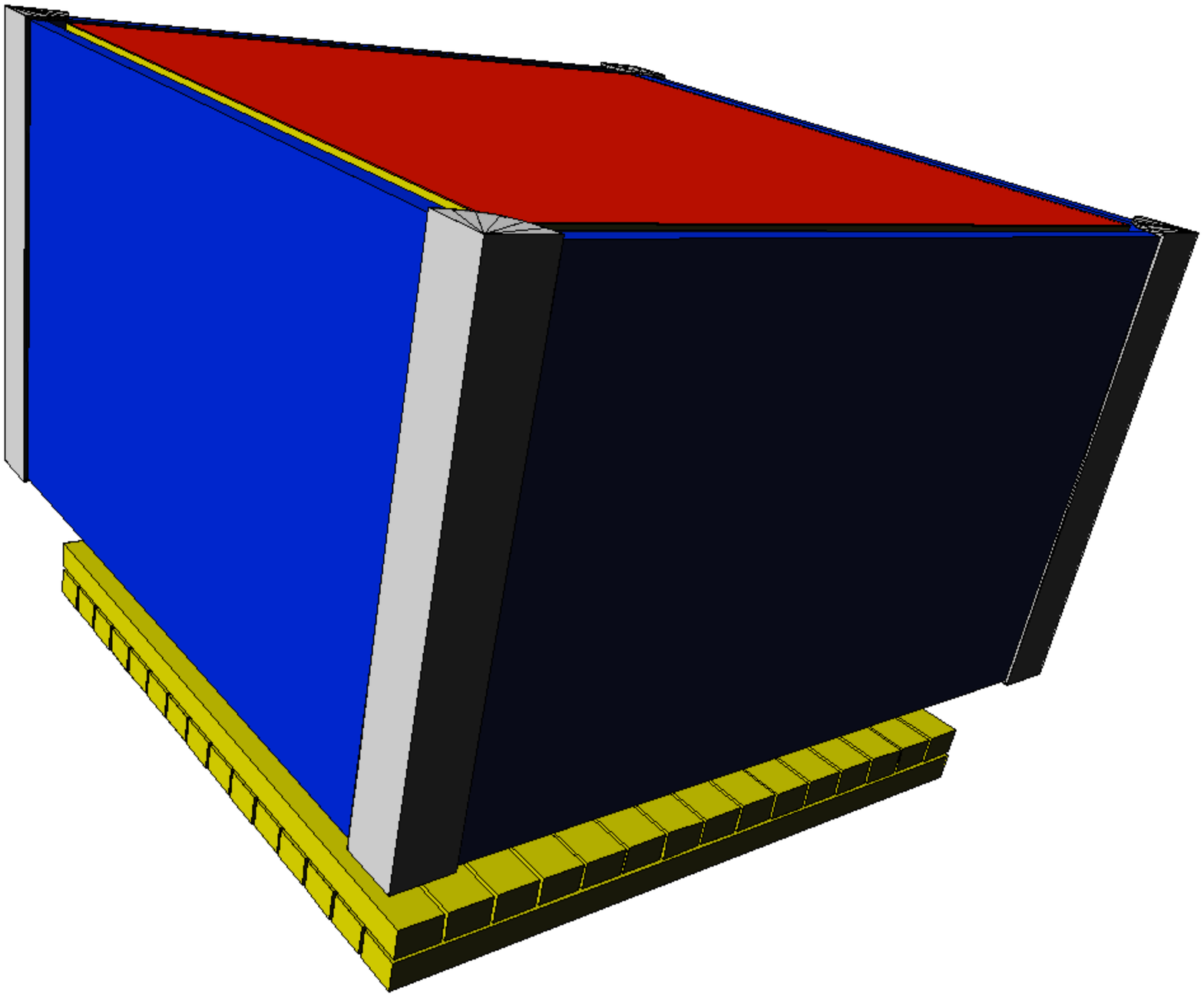}
      \includegraphics[width=0.20\textwidth]{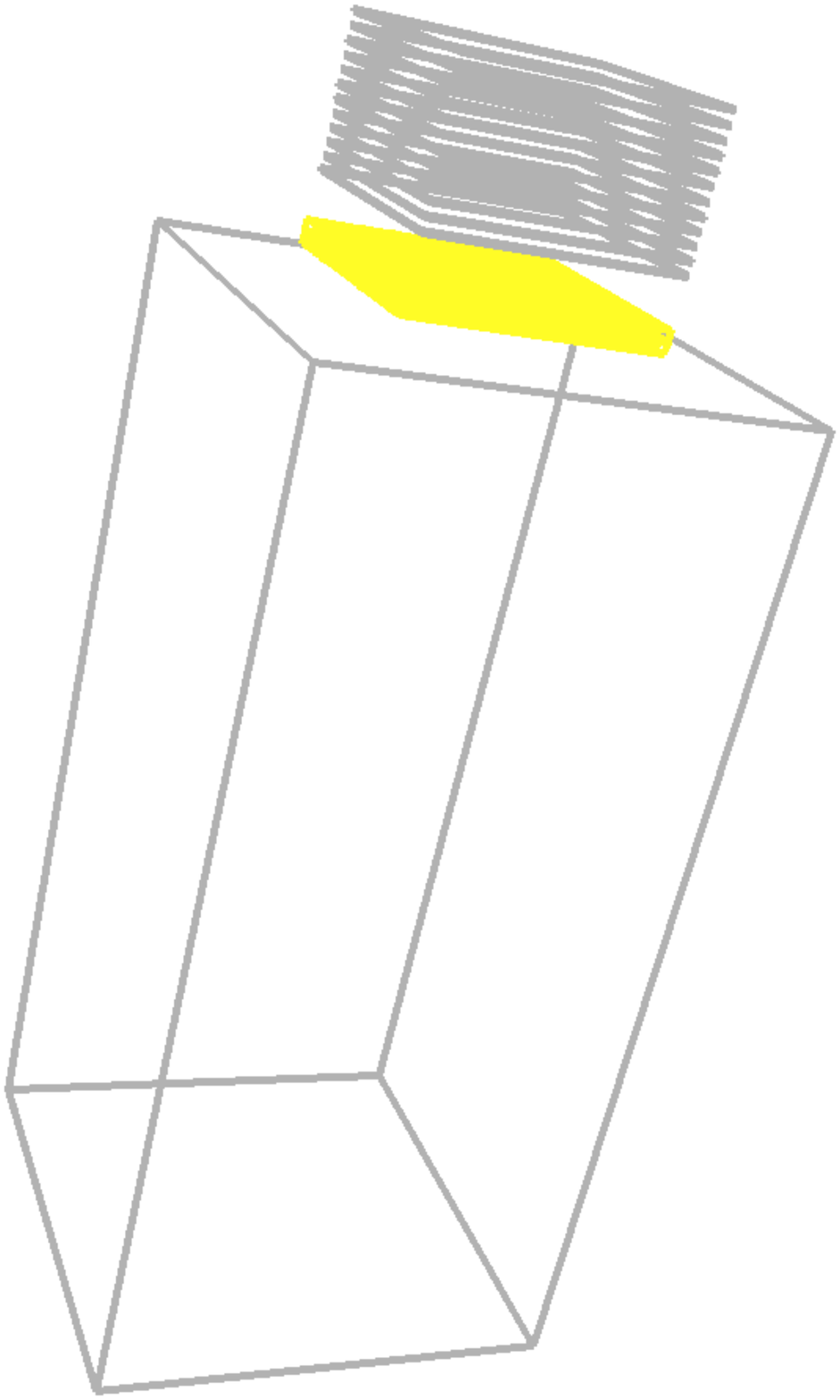}
   \caption{\label{fig:mass_model}\textit{Top row:} orthogonal cut view of the \textit{AGILESim} mass model (left) of the AGILE/GRID, including the tracker, visible in the center, the MCAL, in yellow, at the bottom and the grey ACS surrounding the two instruments. The tracker composition, from the first to the 13th tray, is shown in the center. The X-Y Silicon-strips, in grey, are highlighted in the right figure. \textit{Bottom row:} view of the grey AC panels (left), the blue electronics panels placed at the tracker sides (center), and the wire-framed Aluminum equivalent solid (right) approximating the spacecraft placed under the AGILE payload.}
      \end{figure*}
\section{The \textit{AGILESim} virtual telescope} \label{sec:design}
The Geant4 \citep{g4_1, g4_2, 2016NIMPA.835..186A} toolkit is a Monte Carlo transport code for the simulation of particle interaction with matter.
Initially developed by CERN for the simulation of high energy experiments at particle accelerators, it is now maintained by a large collaboration and has been extended to lower  energies (sub-keV scale).
Geant4 has become the standard tool used by many space agencies (e.g. ESA, NASA) in the radiation shielding analysis and the simulation of the instrument performance (e.g. quantum efficiency, sensitivity) of all major X-ray (e.g., XMM-Newton, ATHENA) and gamma-ray (e.g. AGILE, FERMI) space telescopes. \textit{GAMS}, the former Monte Carlo simulator of the GRID instrument (GEANT AGILE MC Simulator, \citet{2002NIMPA.486..610L, 2002NIMPA.486..623C}), was developed using the FORTRAN library GEANT-3 \citep{Brun1994aa}. The \textit{GAMS} mass model consisted of the spacecraft and the AGILE payload, comprising the GRID instrument, Super-AGILE, the thermal shield, the mechanical structure, and the lateral electronic boards. Used to evaluate the background rejection efficiency and the telescope in-flight calibration \citep{2013A&A...558A..37C}, it has been recently validated against test beam measurements \citep{2018ApJ...861..125C}.
\\
\textit{BoGEMMS} is a Geant4-based simulation framework \citep{2012SPIE.8453E..35B} for the evaluation of the scientific performance (e.g. background spectra, effective area) of high energy experiments, with particular focus on X-ray and gamma-ray space telescopes. Similarly to other Geant4-based frameworks, e.g \textit{MEGAlib} \citep{2006NewAR..50..629Z}, its modularity and parametrized structure allows the user to interactively customize the mass model, the physics processes, and the simulation output at run-time using an external configuration file. \textit{BoGEMMS} records the interactions in FITS\footnote{The common format used by the astronomical community for data interchange and archival storage (https://fits.gsfc.nasa.gov/).} and ROOT\footnote{Analysis and visualization C++ programming library developed by CERN (https://root.cern.ch/).} format output files, then filtered and analysed as a real observation in space. 
In addition to the totally customizable geometry, it provides two application-dependent simulation branches \citep{2014SPIE.9144E..3NF} for the evaluation of (i) the instrumental background of X-ray focusing telescopes and (ii) the scientific performance of gamma-ray electron tracking telescopes.
\\
The gamma-ray branch, conceived as a common, multi-purpose framework for present and future gamma-ray space telescopes, is used here as the core of \textit{AGILESim}, the brand-new AGILE Monte Carlo simulator. The \textit{BoGEMMS} output is analyzed by a first post-processing level that acts as a back-end simulator of the GRID instruments (e.g. digitization of the hits, capacitive coupling) and produces AGILE-like data products to be read by the standard AGILE scientific pipeline. The final output of the analysis pipeline is an event list with associated energy and incoming direction. 
\\
The system of reference used throughout this paper has its origin at the tracker center and the Z-axis directed along the telescope axis. The strips are oriented along the X and Y axis. The photon incident polar angle $\theta$ is the angular distance from the tracker Z-axis, while its azimuthal angle $\phi$ starts from the X-axis. All results are obtained with the Geant4 release 10.4, using the Geant4 electromagnetic \textit{G4EmLivermorePolarizedPhysics} reference physics list, which tracks the interactions of electrons and photons with matter down to about 250 eV using interpolated data tables based on the Livermore library. See Sect. \ref{sec:conf_eff} for a verification of the Geant4 physics library by comparison with tabulated data.

\subsection{Mass model}
Since our goal is to study the reconstruction of pair events, we ignore here the presence of the Super-AGILE instrument, assuming a negligible attenuation efficiency towards gamma-rays. It should be included however for simulations of the particle-induced background level. 
\\
The \textit{AGILESim} three-dimensional mass model is shown in Fig. \ref{fig:mass_model} in all its components: the Silicon tracker and the MCAL (top row), the ACS panels, the electronic boards and the spacecraft equivalent mass (bottom row).

\subsubsection{Silicon tracker}
The GRID ST core device is the Silicon plane, consisting of two separate 410 $\mu$m thick layers of orthogonal single-sided strips to detect the X and Y positions (the lateral \textit{views}) of the electron/positron pair. 
The 12 X-Y Si planes are divided in 3072 physical strips with a pitch of 121 $\mu$m and organized in 13 trays for a total height of $\sim24$ cm. The tray load-bearing structure implementation in \textit{AGILESim} is shown in Fig \ref{fig:mass_model} (top row): $\sim1.5$ cm of Aluminum honeycomb, enclosed by two layers of Carbon fiber. The honeycomb structure is approximated by uniform volumes with Aluminum equivalent density. The Tungsten converter, with a thickness of 245 $\mu m$, is present in the first 10 trays (sky side) and glued on top of the X Si plane, ensuring a total on-axis radiation length of $\sim0.8$ X$_{\rm 0}$.

\subsubsection{MCAL}
The MCAL scintillation calorimeter is consisting of 30 Csi(Tl) bars, $15\times23\times375$ mm$^{3}$ each, divided into two orthogonal sets of 15 rows, for a total radiation length of 1.5 X$_{\rm 0}$ (Fig. \ref{fig:mass_model}, top-right panel). The light produced in the bars is converted into charge by two PIN photodiodes, one at each side of the bar. The longer is the path crossed by the light, the higher is the self-attenuation induced by the scintillation material.
After computing the hit position along the bar, the energy readout by a single diode is emulated by applying experimental attenuation coefficients $\mu$, one for each side, to half the energy deposit in the form:
\begin{equation}
\rm E_{\rm A} = 0.5\times E\times e^{-\mu t} \; ,
\end{equation}
where E is the total deposited energy and \textit{t} is the distance of the energy deposit from the diode.

\subsubsection{ACS, electronic boards and spacecraft}
The ACS, fully surrounding both the ST and the MCAL, is a detection system segmented in 13 panels, 3 on each side and one on the top, constituted by Bicron BC-404 plastic scintillation tiles with a thickness of about 5 mm (Fig. \ref{fig:mass_model}, bottom-left panel). Since plastic scintillators are characterised by poor detection efficiency towards high energy photons, because of the low atomic number of the constituent elements (mainly C and H in this case), if a trigger occurs in one of the panels, it can be assumed to be induced by a charged particle and can be vetoed.
\\
The electronic boards at the tracker sides are simulated by Silicon panels placed inside Aluminum boxes (Fig. \ref{fig:mass_model}, bottom-central panel).
\\
At the bottom of the AGILE GRID system the presence of the spacecraft is emulated by 200 kg of an Aluminum-equivalent box, with a size of $70\times70\times150$ cm$^{3}$ (Fig. \ref{fig:mass_model}, bottom-right panel).
   \begin{figure*}
   \centering
   \includegraphics[trim=0cm 5.cm 0cm 4.cm, clip=true, width=0.99\textwidth]{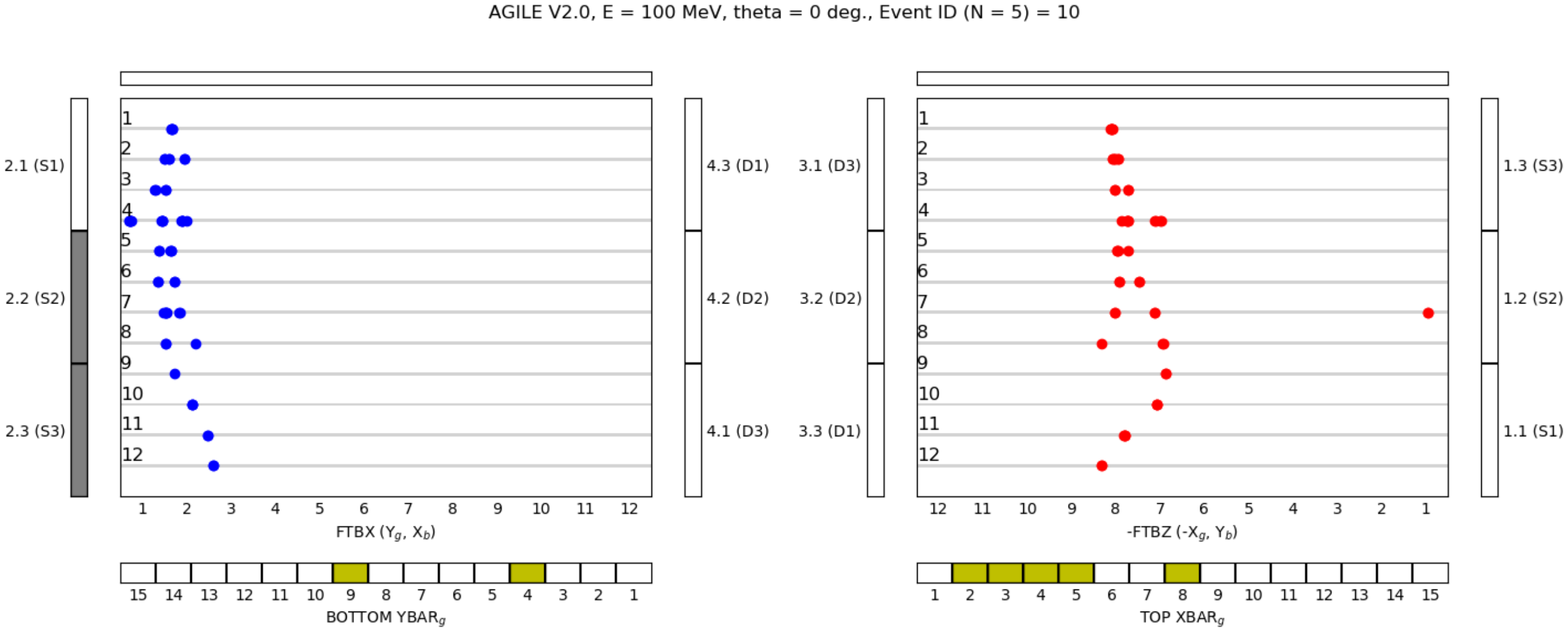}\\
   \includegraphics[trim=3cm 3.cm 4cm 3.cm, clip=true,width=0.57\textwidth]{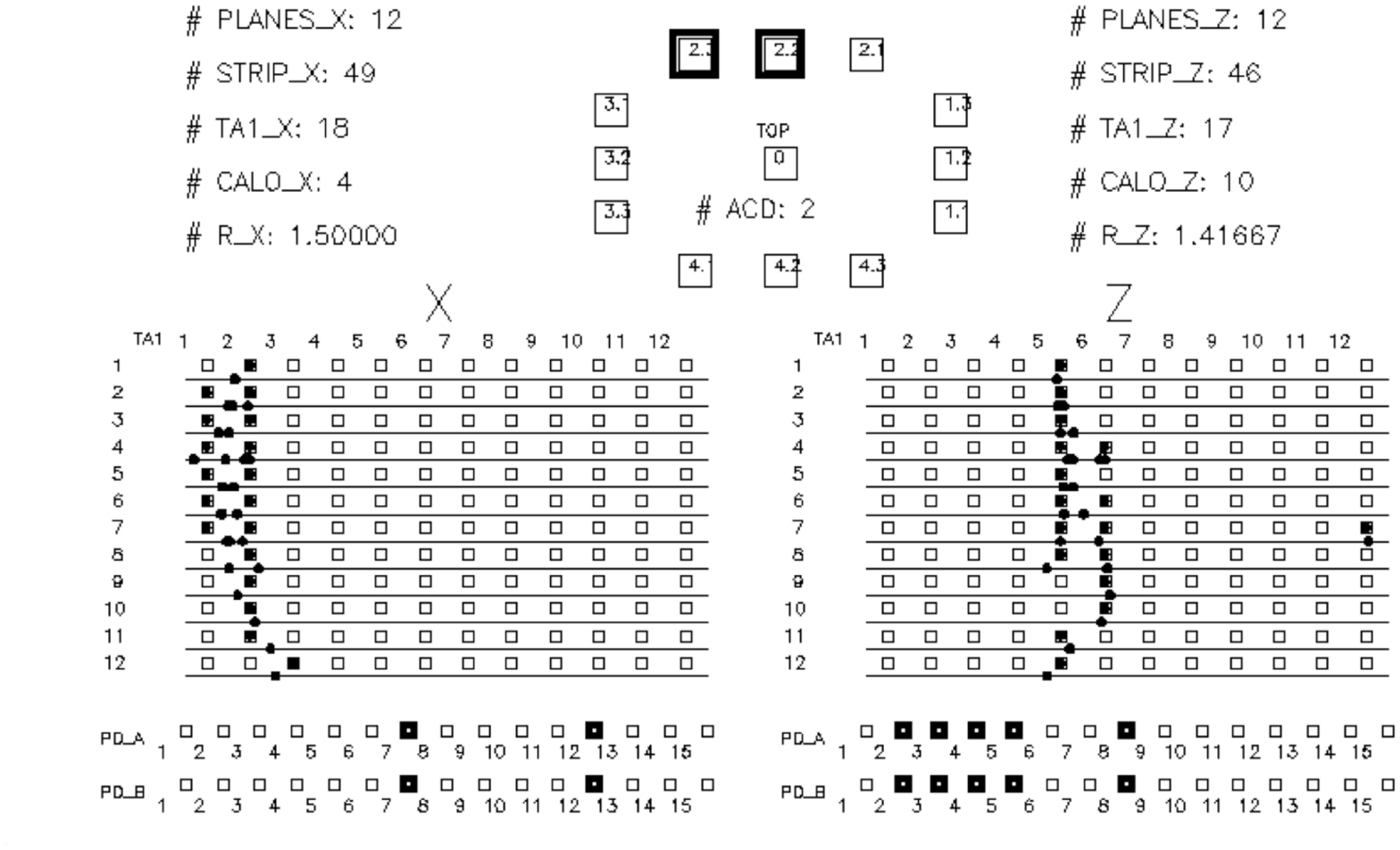}
   \includegraphics[trim=4cm 8.cm 3cm 7.cm, clip=true,width=0.39\textwidth]{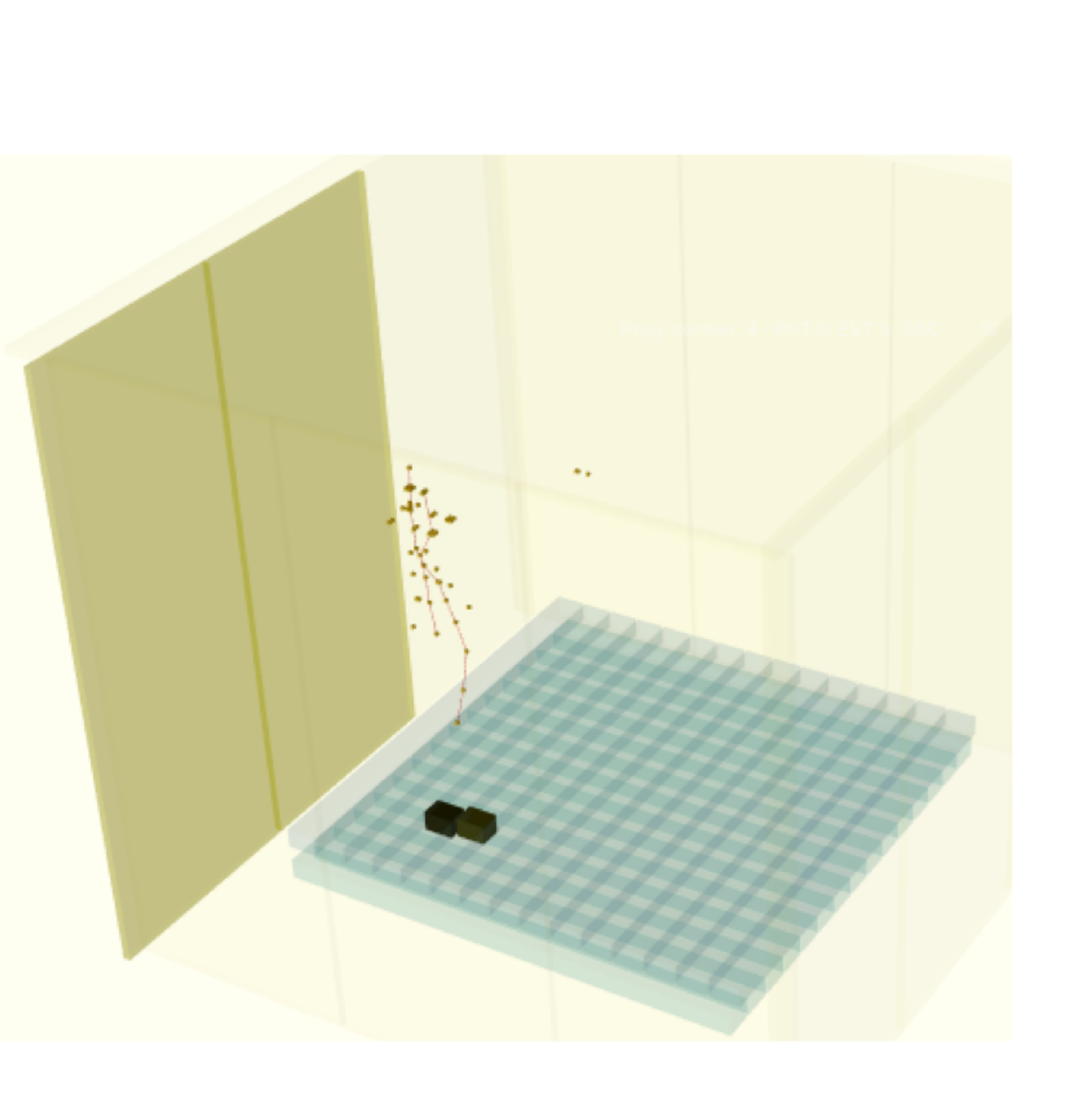}
   \caption{\label{fig:vis}\textit{Top panel:}\textit{AGILESim} visualization of the X and Y lateral view of the ST strips (in blue and red), MCAL bars (in yellow), and fired AC panels (in grey), by a 100 MeV photon induced track. \textit{Bottom panel:} visualization of the same simulated track with the \textit{DHSim} built-in visualization system (left panel) and the AGILE Quick-Look (right panel).}
   \end{figure*}
\subsection{Post-processing}\label{sec:pp}
   \begin{figure}
   \centering
   \includegraphics[trim=0cm 5.cm 0cm 4.cm, clip=true, width=0.48\textwidth]{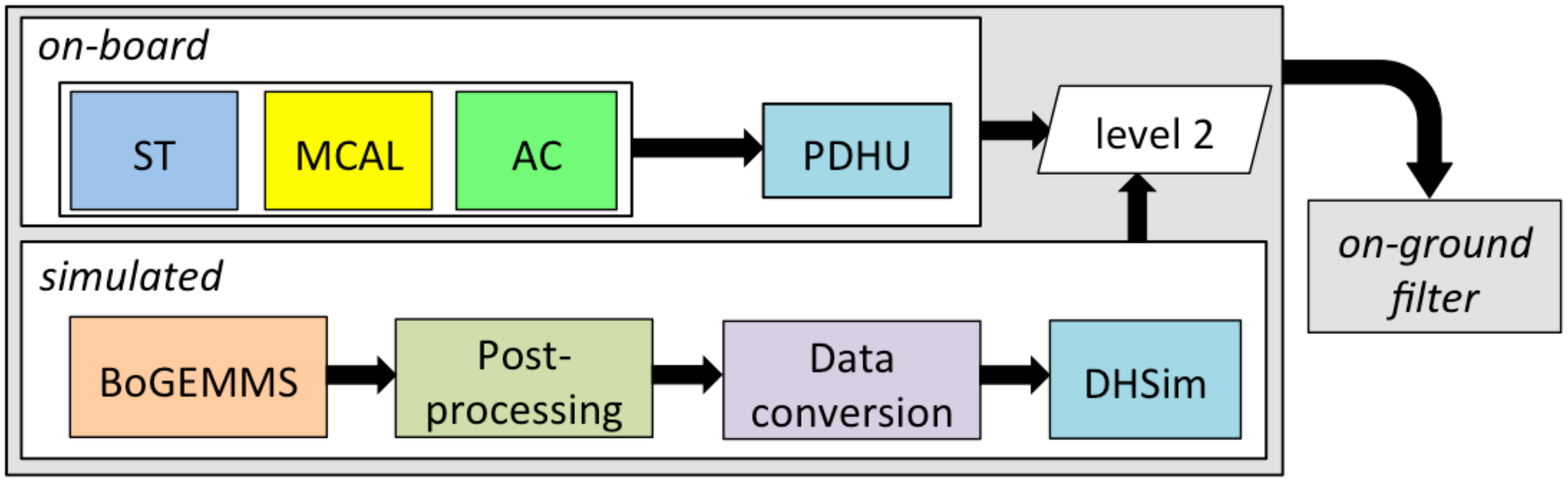}\\
   \caption{\label{fig:proc}Schematic view of the AGILE/GRID on-board processing data flow for both real (on-board) and simulated data.}
   \end{figure}
From the Monte Carlo simulation we directly get the energy deposit in each ST strip, which in the reality is the product of a complex read-out process depending on the design of the electronic devices. The read-out process can be analog (e.g. AGILE), where the true charge is collected from the strips, or digital (e.g. Fermi), where a flag is assigned to the triggered strips. The analog read-out increases the spatial resolution, hence the ability to track the electron/positron pair and consequently the angular resolution.
\\
Including the GRID read-out logic in the simulation requires applying the charge sharing process between adjacent strips during the post-processing of the \textit{BoGEMMS} simulation output. Charge sharing is caused by diffusion during the charge collection and capacitive coupling between the strips.
The first effect is due to diffusion during the drift of the electron-hole pairs along the field lines that increases the size of the charge cloud.
The capacitive coupling is obtained by inserting additional strips between the read-out strips. 
In the ST the read-out pitch, i.e. the distance between two electronic channels connected to the strips, is 242 $\mu$m while the strip pitch, i.e. the physical distance between strips, is 121 $\mu$m. The intermediate strips, or \textit{floating strips}, are placed between contiguous read-out strips. If charge is collected by a \textit{floating} strip, the capacitive coupling induces image charges in the adjacent read-out strips, so that the signal can be measured after applying an interpolation algorithm to the charge distribution. The use of \textit{floating} strips increases the spatial resolution while reducing the number of electronic channels hence the power consumption. The effect of charge sharing in the simulation is obtained by assigning a fraction of the energy deposited in one strip to the neighbouring ones (see \citet{cc1990,Cattaneo:2009eb} for details).
We built for this purpose a tracker look-up table containing only read-out strips and saving in each one an energy amount given by the relation published in \citep{2002NIMPA.486..610L} and obtained from test beam measurements \citep{2002NIMPA.490..146B}. Defining \textit{k} and \textit{i} the read-out and global strip indexes, and their relation \textit{k} = 1 + (\textit{i} - 1)/2, the total energy deposited in the \textit{k}-th read-out strip is E$_{k}$ = E$_{i}$ + 0.38$\cdot$(E$_{i-1}$ + E$_{i+1}$) + 0.115$\cdot$(E$_{i-2}$ + E$_{i+2}$) + 0.095$\cdot$(E$_{i-3}$ + E$_{i+3}$) + 0.045$\cdot$(E$_{i-4}$ + E$_{i+4}$) + 0.035$\cdot$(E$_{i-5}$ + E$_{i+5}$).
\\
The instrumental noise rms of the GRID strips is of the order of 30 ADC counts \citep{2010NIMPA.614..213B}. Given a conversion factor of 0.174 keV/ADC counts, a $\sigma \sim 5$ keV is added to each strip, although it has negligible effects on the trigger rates \citep{2002NIMPA.486..610L}.
After summing all energy deposits within the same strip, generated by the same primary simulated photons, we apply a trigger threshold of 1/4 the energy of a minimum ionising particle (MIP, $\sim27$ keV) to the ST strips \citep{2002NIMPA.490..146B}, as in the real case. 

\subsection{The interface to the AGILE processing pipeline}
In the simulation of AGILE observations, the correct reconstruction of the conversion process of gamma-rays to e$^{-}$/e$^{-}$ pairs is not enough.
Being an indirect detection process, obtaining the primary gamma-ray energy and direction requires a trigger and filtering algorithm to be applied to the Monte Carlo simulation. We make use of the AGILE/GRID standard processing pipeline \citep{2019ExA....48..199B}, which takes as input the trigger list from the instruments (ST, MCAL and AC) and produces at the end a classified event list for the scientific analysis (e.g. production of counts map, spectra and light curves of the science target). 
\\
The first block of the pipeline is the on-board Payload Data Handling Unit (PDHU, \citet{dhsim, 2019RLSFN.tmp...27A}) that applies the hardware and software veto logic to the scientific data and produces the \textit{level 2} telemetry data format to be transmitted to the ground station. When Monte Carlo simulations are used as input, the Data Handling Simulator (\textit{DHSim}) first converts the Monte Carlo data into the instrument trigger data format and then acts as a virtual Data Handling Unit, producing the telemetry files for the on-ground filtering pipeline. In order to use the \textit{DHSim} software, we converted the \textit{AGILESim} reference system and data output to the same format used in \textit{GAMS}, including the X and Y views of the ST with the triggered strips in MIP units, the fired AC panels and the energy deposited in the MCAL bars.
\\
The on-ground \textit{FM3.119} filter is then used for the track identification and reconstruction, that first selects the potential tracks of the pair \citep{2002NIMPA.488..295P} and then fits the track parameters using a recursive Kalman filter algorithm \citep{2006NIMPA.568..692G}. The track minimizing the sum of the $\chi^{2}$ of the ST planes is associated to the primary gamma-ray conversion. 
\\
We developed an independent GRID event visualizer to directly display the BoGEMMS tracks and compare them to the tracks loaded by the on-ground filter. 
In Fig. \ref{fig:vis} (top panel) the \textit{AGILESim} GRID visualizer displays as example a 100 MeV gamma-ray with an incoming direction parallel to the telescope axis. The same event is displayed in Fig. \ref{fig:vis} (bottom panels) by the \textit{DHSim} and the AGILE filter Quick-Look software \citep{2008SPIE.7011E..3CB}.  Fig. \ref{fig:proc} shows a simplified schema of the processing pipeline from the on-board or simulated data read-out to the on-ground filter.
  \begin{figure}
   \centering
   \includegraphics[trim=3cm 1.cm 3cm 1.cm, clip=true,width=0.47\textwidth]{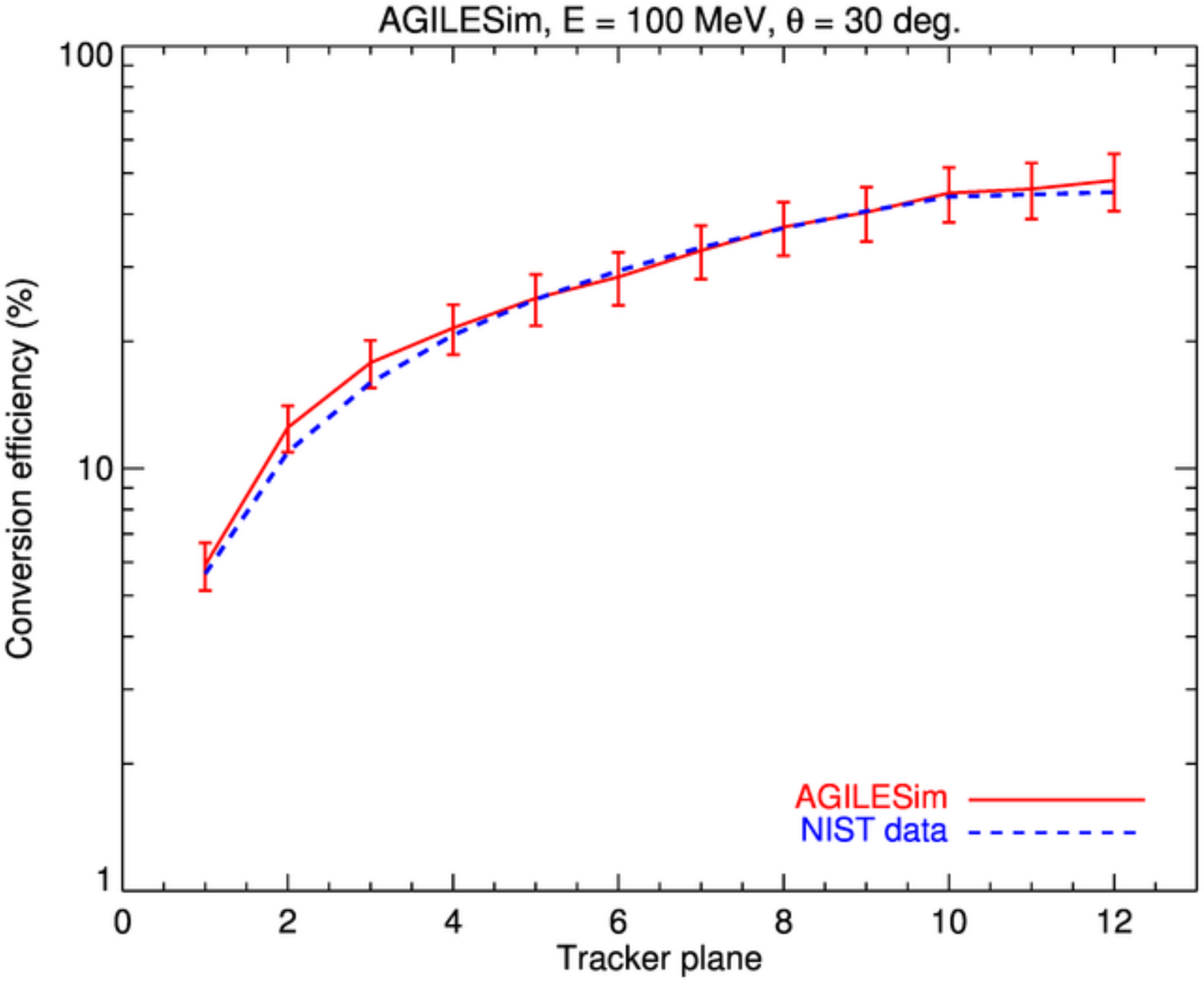}
   \caption{\label{fig:conv}Pair conversion efficiency of a 100 MeV mono-energetic beam, impinging with an off-axis angle of 30$^{\circ}$, as a function of the tracker plane (numbered starting from the sky-side).}
   \end{figure}
     \begin{figure*}
   \centering
   \includegraphics[width=0.49\textwidth]{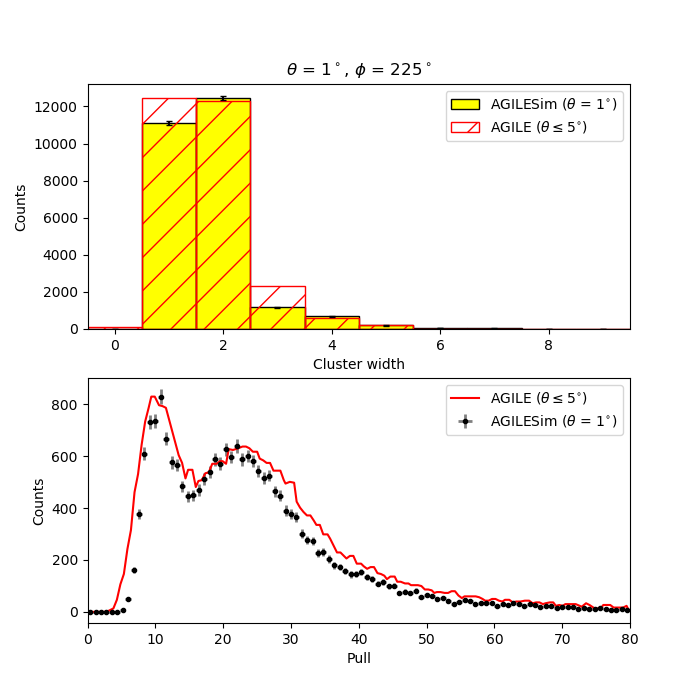}
      \includegraphics[width=0.49\textwidth]{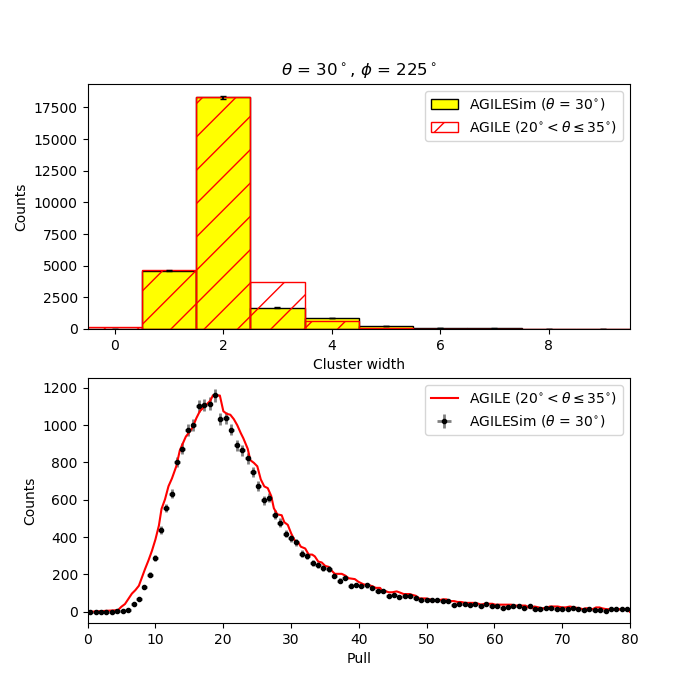}
   \caption{\label{fig:pull}Simulated distribution of cluster width (top panels) and pull (bottom panels) for $\theta=1^{\circ}$ (left panels) and $\theta=30^{\circ}$ (right panels) inclination angles. The measured distribution obtained during the on-ground AIV tests \citep{2010NIMPA.614..213B}, for $\theta \leq 5^{\circ}$ and $20^{\circ} <\theta\leq35^{\circ}$, is shown in red.}
   \end{figure*}
\section{Scientific verification and validation}\label{sec:validation}

The physics reliability of \textit{AGILESim} in reproducing the AGILE/GRID performance is demonstrated by comparisons with (i) the analytical  evaluation of the tracker conversion efficiency from tabulated data, (ii) on-ground calibration measurements, and (iii) the in-flight angular resolution from observations of the Vela pulsar.  

\subsection{Conversion efficiency}\label{sec:conf_eff}
The ST conversion efficiency, i.e. the probability of converting a gamma-ray into an electron/positron pair, depends, besides on the input photon energy and direction, on the tracker design. We evaluate this efficiency in \textit{AGILESim} by simulating mono-energetic parallel photon beams with energy 50, 100, 200, 400 and 1000 MeV and with an off-axis angle $\theta = 30^{\circ}$, then computing the ratio between the gamma-rays producing a pair and the ones not interacting (Compton scatterings can be neglected at this regime), as a function of the number of crossed planes. The pair production attenuation coefficients \textit{$\mu$} of the material composing the ST trays  are tabulated by the National Institute of Standards and Technology (NIST, \citet{nist}). They are used in the Beer-Lambert formula to compute the expected conversion efficiency for each layer of thickness \textit{t}, in the form $(1 - e^{-\mu  t / \rm cos(\theta)})$. 
The resulting simulated conversion efficiency is shown in blue in Fig. \ref{fig:conv} for an input energy of 100 MeV and compared with the corresponding analytical values, plotted in red. The error bars are Poisson statistical fluctuations from the number of simulated photons. The simulation reproduces quite well the theoretical efficiency based on the material cross-section. This comparison verifies not only the correct building of the \textit{AGILESim} ST mass model but also the Geant4 pair production physics models implemented in the \textit{BoGEMMS} simulation framework.

\subsection{On-ground verification}\label{sec:aiv}
Testing the correct simulation of electrons and positrons interacting with the Si strip detectors is another fundamental step of the physics verification. In the AGILE AIV (Assembly, Integration, and Verification) on-ground activity \citep{2010NIMPA.614..213B}, the interaction of atmospheric muons was used to characterize the ST noise and front-end read-out. 
Atmospheric muons behave as MIPs, i.e. they deposit in the Si layers a minimum amount of energy per path unit that is almost independent of the incident energy. The electrons and positrons interacting with the ST strips behave as MIP as well, and we can use the cluster charge distribution collected in the AIV stage to verify the \textit{AGILESim} simulation.
\\
A cluster of strips is a set of contiguous read-out strips triggered by the passage of a particle and its identification is performed on board by the \textit{PDHU}. The ST \textit{pull} is defined as the ratio between the maximum signal in the cluster strips and the instrumental noise rms, having a mean value of 5 keV (see Sec. \ref{sec:pp}). The number of strips populating the clusters, defined as cluster width, and their \textit{pull} were measured during the GRID on-ground noise tests for a set of inclination angles of atmospheric muons with respect to the GRID axis. In order to compare such distributions with the \textit{AGILESim} output, we simulated the on-ground atmospheric muons as beams of parallel particles hitting on the tracker surface at two inclination values ($\theta = 1^{\circ}$ and $30^{\circ}$) and an averaged azimuthal angle of 45$^\circ$ ($225^{\circ}$ in the BoGEMMS system of reference). According to \cite{2018IJMPA..3350175S}, the on-ground spectral distribution $I(E)$ of muons with momentum in the range from 1 GeV/c to 1 TeV/c can be modelled as:
\begin{equation}
    \rm I(E) = I_{0}\;N\;(E_{0} + E)^{\rm -n}\left(1 + \frac{\rm E}{\epsilon}\right)^{-1}\;,
\end{equation}
where $E$ is the kinetic energy of the muon expressed in GeV, $I_{0} = 72.5$ m$^{-2}$ s$^{-1}$ sr$^{-1}$, $n$ = 3.06, $E_{0}= 3.87$ GeV and $\epsilon=854$ GeV at sea-level\footnote{The AIV measurements performed in Milan, with an altitude $\sim100$ m, are considered for simplicity at sea-level.}.
\begin{figure*}
   \centering
  \includegraphics[width=0.48\textwidth]{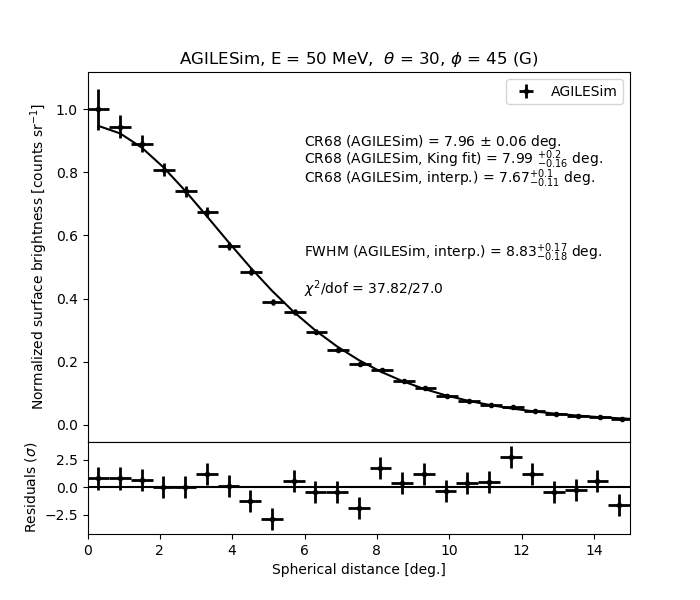}
  \includegraphics[width=0.48\textwidth]{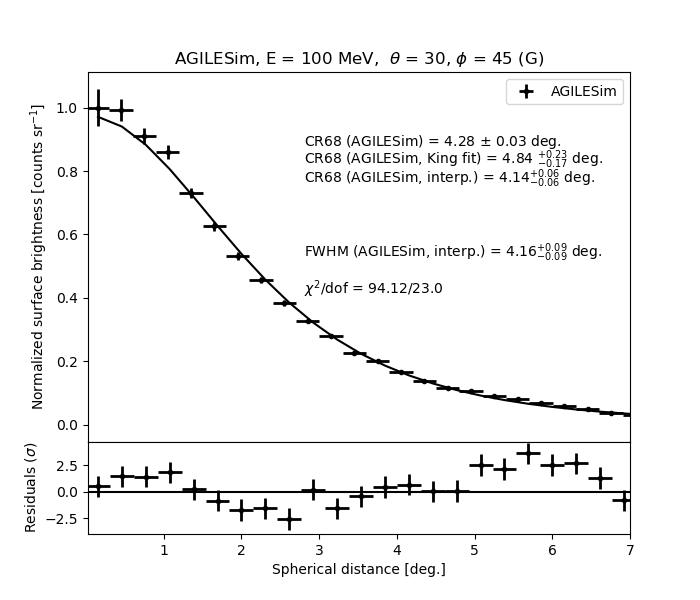}\\
  \includegraphics[width=0.48\textwidth]{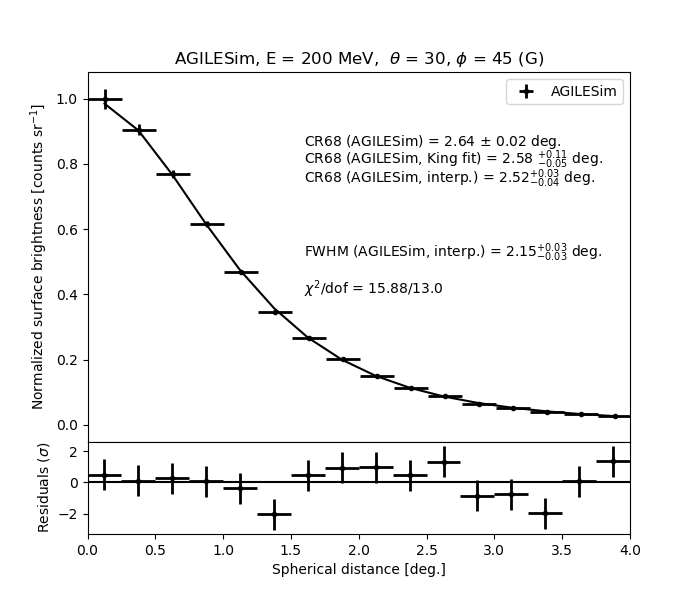}
  \includegraphics[width=0.48\textwidth]{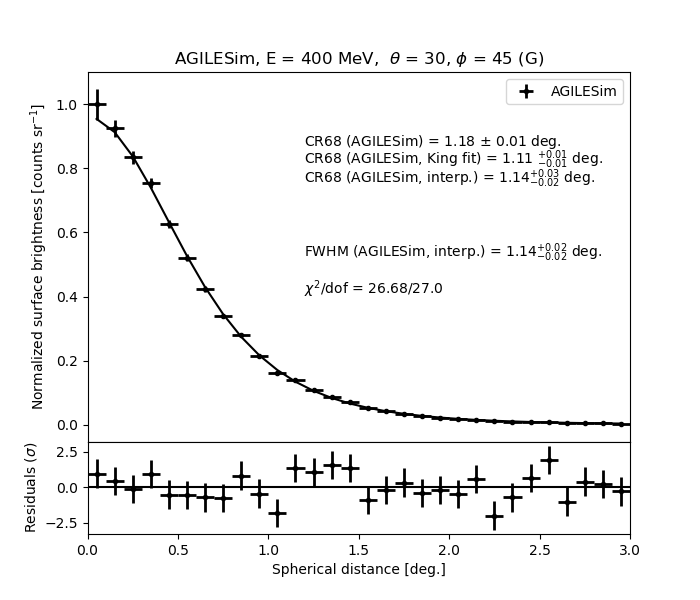}\\
  \includegraphics[width=0.48\textwidth]{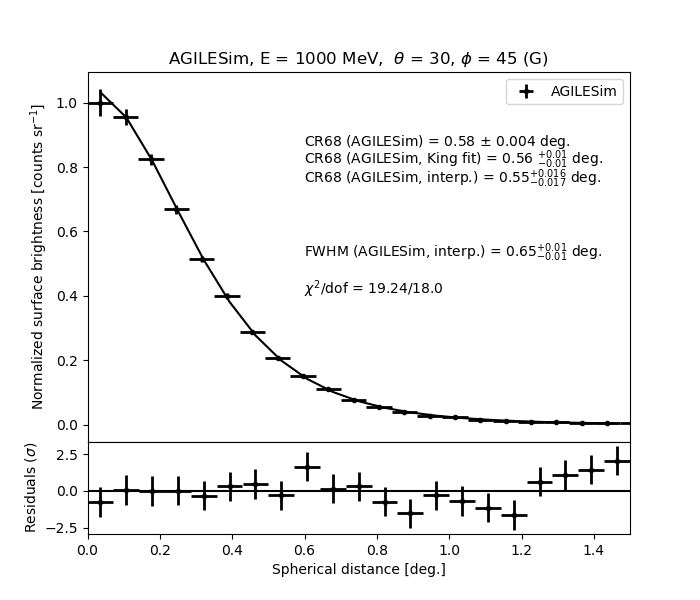}
  \includegraphics[width=0.46\textwidth]{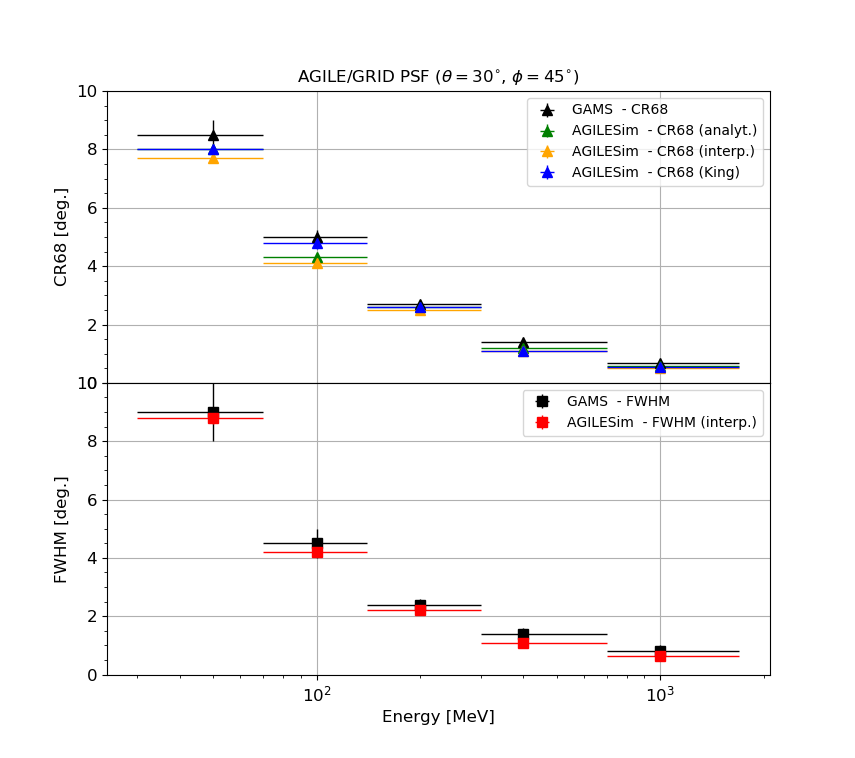}
   \caption{\label{fig:simpsf} Simulated PSF radial profile, in surface brightness, as a function of the spherical distance from the true direction for a set of mono-energetic planar sources at 50, 100, 200, 400, and 1000 MeV. The resulting CR68 and FWHM PSF values are compared to published \textit{GAMS} simulations \citep{2015ApJ...809...60S}.} 
 \end{figure*}
\\
The resulting cluster width and \textit{pull} distributions are shown in Fig. \ref{fig:pull} (top and bottom panels respectively) and compared to the real on-ground measurement \citep{2010NIMPA.614..213B} at similar muon incident angles. 
\\
For large incidence angles, the probability for a muon to cross at the same time both \textit{reading} and \textit{floating} strips is quite high. For this reason, at $\theta \sim 30^{\circ}$ most of the clusters are composed by two strips, whereas at inclination near to zero the probability of having single strip or double-strip clusters is almost the same. For both inclinations, the simulated cluster width distributions reproduce very well the real ones.
\\
When a muon hits a \textit{floating} strip, the signal is given by the charge induced on the adjacent \textit{readout} strips, and its amplitude is lower because only a fraction of the generated charge can contribute to the signal. On the contrary, when a muon hits a \textit{readout} strip, the signal amplitude is given by the full amount of produced charge.
These two mechanisms determine the shape of the \textit{pull} distribution. Indeed, for a small incidence angle, when \textit{floating} and \textit{readout} strips are most likely individually hit, the \textit{pull} distribution shows two separated peaks, where the one at lower \textit{pull} values is due to the \textit{floating} strips contribution. On the contrary, when the incidence angle is higher, each muon can affect both the \textit{floating} and \textit{readout} strips, and the two peaks merge into a single peak. In both cases, the simulated response reproduces quite well the real data.

\subsection{PSF for mono-energetic sources}\label{sec:space}
The angular resolution of a gamma-ray telescope defines its ability to discriminate between two point sources placed in proximity to each other. Besides avoiding source confusion in crowded sky regions (e.g., the galactic center), the angular resolution is a key factor affecting the instrument performance, because the smaller is the imaged shape of the point source, the lower is the background contribution within, hence the better the overall sensitivity to detect faint sources. The PSF is a probability distribution for photons from a point source being reconstructed at a certain spherical distance from their true direction. Assuming azimuthal symmetry of such function, as it is observed in gamma-ray telescopes below an off-axis angle of 30$^{\circ}$ \citep{2015ApJ...809...60S}, we describe the PSF in terms of surface brightness, i.e. counts per solid angle, as a function of the spherical distance from  the true direction. The output of the AGILE filter and reconstruction pipeline gives, for a certain number of simulated gamma-rays at ($\theta$, $\phi$) = (30$^{\circ}$, 45$^{\circ}$) and an input true energy $E$, a list of events with reconstructed direction $\theta_{\rm rec}$ and $\phi_{\rm rec}$ and reconstructed energy $E_{\rm rec}$. A $G$ flag labels all the events reconstructed as a likely gamma-ray \citep{2013A&A...558A..37C}. All present results refer to \textit{G-flagged} events, as those used in the in-flight scientific analysis. We simulate parallel beams of photons hitting the full surface of the tracker, each with a constant energy of 50, 100, 200, 400 and 1000 MeV. For our analysis we select all the photons with a reconstructed energy within defined energy bands (30 -- 70 MeV, 70 -- 140 MeV, 140 -- 300 MeV, 300 -- 700 MeV and 700 -- 1700 MeV) to compare BoGEMMS with the GAMS simulations reported in \citet{2015ApJ...809...60S}.
\\
The spherical distance \textit{d$_{\rm sph}$} is given by:
\begin{eqnarray}
  d_{\rm sph} & = & \rm cos^{-1}(sin(\theta_{rec})sin(\theta)cos(\phi_{rec} - \phi) + \\
  & & + \rm cos(\theta_{rec})\rm cos(\theta)) \; .
   \end{eqnarray}
The photon true direction ($\theta$, $\phi$) is assumed to be the center of a new system of reference, and each reconstructed photon is characterized by its radial distance from the center. The spherical distance from the true direction is then equally binned in annular regions around the center, subtending an increasing solid angle from the center. The number of counts in each region is divided by its subtended solid angle to compute the surface brightness, and then the distribution maximum is normalized to 1. The uncertainty for each bin is given by the Poisson statistics. The simulated PSF radial profiles at 50, 100, 200, 400 and 1000 MeV are shown in Fig. \ref{fig:simpsf}.
  \begin{table*}
\begin{center}
\begin{small}
\begin{tabular}{c|c|c|c|c|c|c|c}
\multicolumn{8}{c}{\textsc{PSF - Mono-energetic Source ($\theta=30^{\circ}$, $\phi=45^{\circ}$)}}\\
\hline
\hline
\multirow{5}{*}{} & \multicolumn{5}{c|}{\multirow{2}{*}{AGILESim (G--class)}} & \multicolumn{2}{c}{\multirow{2}{*}{GAMS (G--class)}} \\
 & \multicolumn{5}{c|}{} & \multicolumn{2}{c}{}\\
\cline{2-8}
Energy & \multicolumn{1}{c|}{\multirow{2}{*}{analytical}}& \multicolumn{2}{c|}{\multirow{2}{*}{interpolation}} & \multicolumn{2}{c|}{\multirow{2}{*}{King fit}} & \multicolumn{2}{c}{\multirow{2}{*}{analytical}} \\
& \multicolumn{1}{c|}{}& \multicolumn{2}{c|}{}& \multicolumn{2}{c|}{} & \multicolumn{2}{c}{} \\
\cline{2-8}
[MeV] & \multirow{2}{*}{CR68}& \multirow{2}{*}{CR68} & \multirow{2}{*}{FWHM}& \multirow{2}{*}{CR68}& \multirow{3}{*}{$\chi^{2}/\rm dof$}& \multirow{2}{*}{CR68} & \multirow{2}{*}{FWHM}\\
& & & & & & & \\
& [$^{\circ}$] & [$^{\circ}$] & [$^{\circ}$]& [$^{\circ}$]& &  [$^{\circ}$]& [$^{\circ}$] \\
\hline
\multirow{2}{*}{50}&  \multirow{3}{*}{$8.0 \pm 0.1$}  & \multirow{3}{*}{$7.7_{-0.1}^{+0.1}$}&  \multirow{3}{*}{$8.8_{-0.2}^{+0.2}$} & \multirow{3}{*}{$8.0_{-0.2}^{+0.2}$}& \multirow{3}{*}{38/27}& \multirow{3}{*}{$8.5\pm0.5$}& \multirow{3}{*}{$9.0\pm1.0$}\\
& & & & & & &\\
(30 - 70)&  & & & & & & \\
\multirow{2}{*}{100}&  \multirow{3}{*}{$4.3 \pm 0.03$} & \multirow{3}{*}{$4.1_{-0.1}^{+0.1}$}&  \multirow{3}{*}{$4.2_{-0.1}^{+0.1}$} & \multirow{3}{*}{$4.8_{-0.2}^{+0.2}$}& \multirow{3}{*}{94/23}& \multirow{3}{*}{$5.0\pm0.25$}&\multirow{3}{*}{$4.5\pm0.5$}\\
& & &  & & &  &\\
(70 - 140)& & & & & & &\\
\multirow{2}{*}{200}&  \multirow{3}{*}{$ 2.6\pm0.02 $}  & \multirow{3}{*}{$2.5_{-0.03}^{+0.04}$}&  \multirow{3}{*}{$2.2_{-0.03}^{+0.03}$} & \multirow{3}{*}{$2.6_{-0.1}^{+0.1}$}& \multirow{3}{*}{16/13}& \multirow{3}{*}{$2.7\pm0.1$}&\multirow{3}{*}{$2.4\pm0.2$} \\
& & & & &  & &\\
(140 - 300)& & & & &  & &\\
\multirow{2}{*}{400}&  \multirow{3}{*}{$1.2\pm 0.01$} & \multirow{3}{*}{$1.1_{-0.03}^{+0.02}$}&  \multirow{3}{*}{$1.1_{-0.02}^{+0.02}$}  & \multirow{3}{*}{$1.1_{-0.01}^{+0.01}$}& \multirow{3}{*}{27/27}& \multirow{3}{*}{$1.4\pm0.1$}& \multirow{3}{*}{$1.4\pm0.2$}\\
& & & & &  & &\\
(300  - 700)& & & & & & &\\
\multirow{2}{*}{1000}&  \multirow{3}{*}{$0.58 \pm 0.004$}  & \multirow{3}{*}{$0.51_{-0.004}^{+0.004}$}&  \multirow{3}{*}{$ 0.65_{-0.01}^{+0.01} $}  & \multirow{3}{*}{$0.56_{-0.01}^{+0.01}$}& \multirow{3}{*}{19/18}& \multirow{3}{*}{$0.7\pm0.1$}&\multirow{3}{*}{$0.8\pm0.2$}\\
& & & & & & &\\
(700  - 1700)& & & & & & &\\
\hline
\end{tabular}
\end{small}
\end{center}
\caption{\label{tab:mono_psf} Comparison of the values of PSF for AGILE/GRID obtained simulating a mono-energetic source with \textit{AGILEsim} and \textit{GAMS} \citep{2015ApJ...809...60S}}.
\end{table*}
In gamma-ray astronomy, the angular resolution is described as the radius containing a certain percentage of photons in the PSF, usually the 68\%, or the full width half maximum (\textit{FWHM}) of its radial profile. The containment radius (\textit{CR68} from now on) takes into account the whole distribution, including the length of the tail. The FWHM instead measures the central shape of the profile. For these reasons, the PSF \textit{CR68} and \textit{FWHM} can have different values and a different dependence on the source energy and off-axis angle.
\\
Both parameters can be computed analytically, by interpolation, or by fitting the PSF with an appropriate function. 
We obtain the analytical 68\% containment radius by just counting the reconstructed photons: in this case the 1$\sigma$ error is given by the statistics of the simulation in terms of Poisson uncertainty $\rm \sqrt N$.
The normalized radial profile can be linearly interpolated by a function and the spherical distance at which the function is 0.5 defines the
half width of the half maximum. Twice this value gives the \textit{FWHM}. For each bin of the radial profile, the $1 \sigma$ lower and the upper limits are also interpolated in the same way to extract the corresponding \textit{FWHM} upper and lower limits. We also report the \textit{CR68} obtained from the interpolation function, with errors computed from the $1 \sigma$ lower and upper limits as the  \textit{FWHM}.
\\
The PSF radial profile of a point source is often best described not by a Gaussian profile but a superimposition of an exponential core and an $r^{2}$ tail, with \textit{r} indicating the spherical or radial distance. This function, presented in \citet{1971PASP...83..377K} and referred as \textit{King} in \citet{2011A&A...534A..34R} for the evaluation of XMM-Newton angular resolution, takes the form:
\begin{equation}
\rm K(r, \sigma, \gamma) = \frac{1}{2\pi\sigma^{2}}\left(1 - \frac{1}{\gamma}\right)\left(1 + \frac{r^{2}}{2\sigma^{2}\gamma}\right)^{-\gamma}\;,
\end{equation}
with $\sigma$ and $\gamma$ characterizing the size of the angular distribution and the weight of the tail respectively. 
In the small angle approximation $\sin(r)\simeq r$, the King profile integrated over the radial distance is normalized to 1, as a probability distribution. Since we normalize the PSF to its maximum, a multiplicative normalization parameter is added to the King function. 
The AGILE/GRID calibration \citep{2013A&A...558A..37C} used a single King function to model the PSF radial profile from Monte Carlo simulations and  produce the mission instrument response function.  We decided to follow the same procedures of the mission calibration team, and optimization studies of the AGILE scientific performance will be the subject of future works on this subject.
\\
The simulated PSF radial profile is fitted by the King profile, and the resulting $\sigma$ and $\gamma$ parameters are used for the evaluation of the \textit{CR68} radius as derived by \citet{2018ApJ...861..125C}:
\begin{equation}
\rm CR68 = \sigma\sqrt{2\gamma\left((1 - 0.68)^{\frac{1}{1-\gamma}}-1\right)} \;.
\end{equation}
The $1\sigma$ error on the CR68 is obtained from the 68\% confidence contour levels of the $\sigma$ and $\gamma$ best fit values. 
The best fit King models are superimposed to the radial profiles in Fig. \ref{fig:simpsf}, and the residuals, in standard deviations, are plotted in the bottom panels. We are able to describe all mono-energetic profiles with a single King profile with residuals below a $3\sigma$ deviation from the model.
\begin{figure*}
   \centering
   \includegraphics[width=0.49\textwidth]{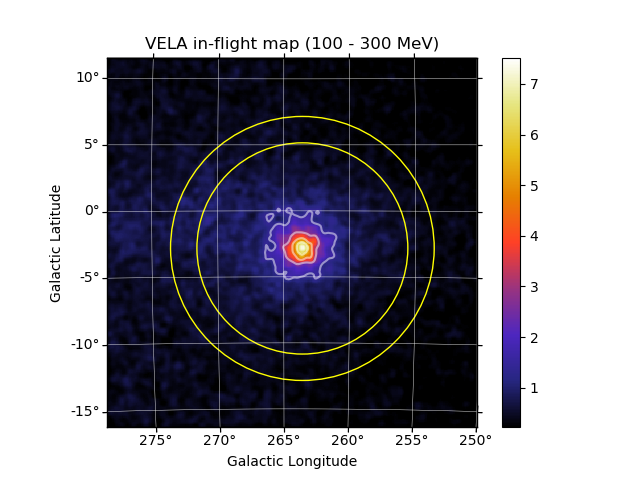}
   \includegraphics[width=0.49\textwidth]{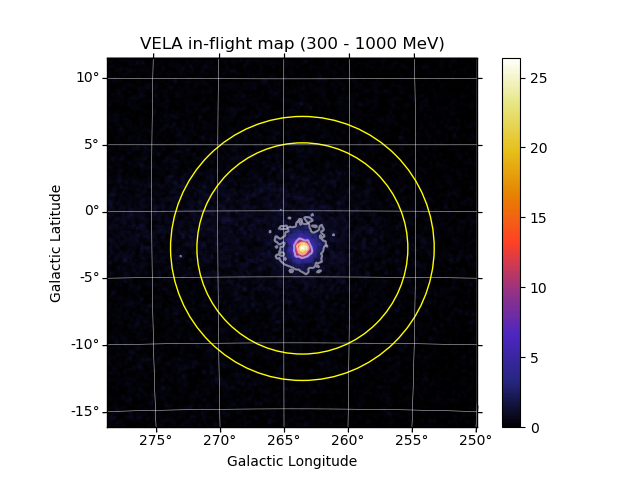}
   \caption{\label{fig:vela_map_real} The Vela pulsar count maps detected by AGILE/GRID in the 100 -- 300 MeV (left panel) and the 300 -- 1000 MeV (right panel) energy range at a $20^{\circ}<\theta<40^{\circ}$ off-axis angle. The yellow rings refer to the background extraction region at $8^{\circ}-10^{\circ}$ radial distance from the source.}
   \end{figure*}
   \begin{figure*}
   \centering
   \includegraphics[width=0.49\textwidth]{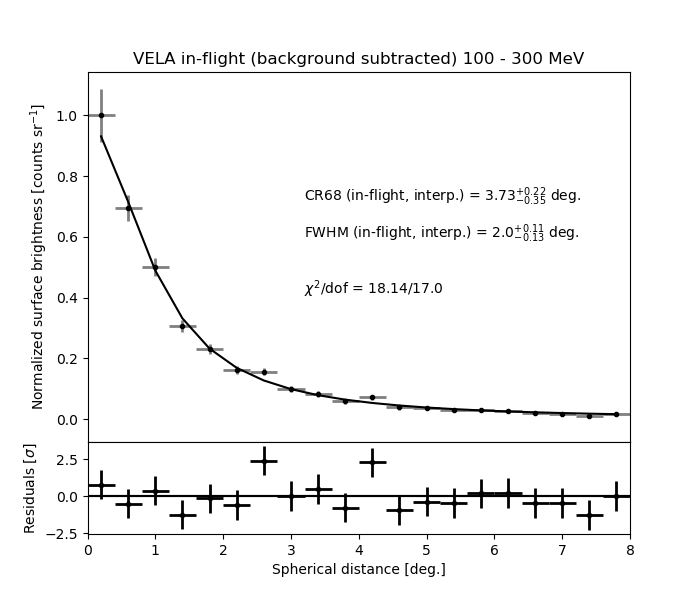}
   \includegraphics[width=0.49\textwidth]{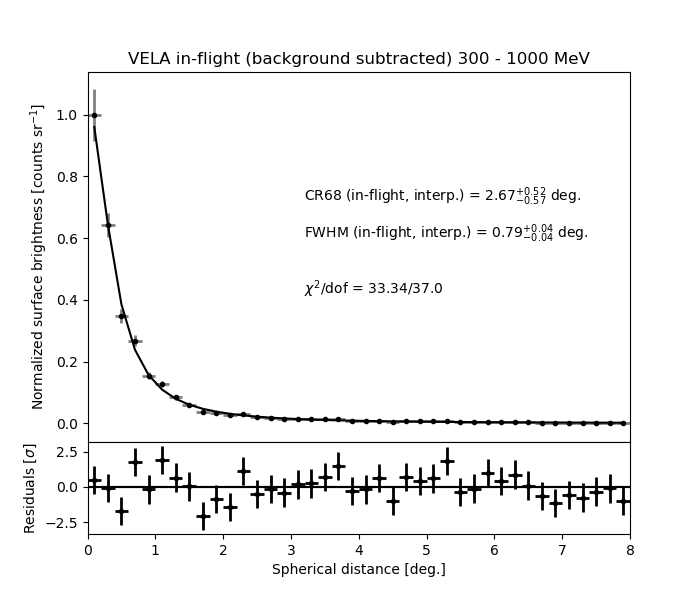}
   \caption{\label{fig:vela_psf_real} The background subtracted in-flight Vela pulsar radial profile in the 100 -- 300 MeV (left panel) and the 300 -- 1000 MeV (right panel) energy bands. The profile is fitted by a single King function (black line) and the residuals, in standard deviations, are plotted in the bottom panel.}
   \end{figure*}
\\
The resulting \textit{AGILESim} PSF obtained for the two methods, \textit{CR68} and \textit{FWHM}, and three different procedures (analytical, interpolation and fitting), are listed in Table \ref{tab:mono_psf}, which also reports for comparison the angular resolution obtained by the \textit{GAMS} simulator in the same energy channels \citep{2015ApJ...809...60S}. The \textit{CR68} and \textit{FWHM} values from \textit{GAMS} are obtained from analytical procedures, and the reported errors derived from the binning applied to the radial profile. The last panel in Fig. \ref{fig:simpsf} summarizes the tabulated angular resolutions obtained with \textit{AGILESim} and \textit{GAMS} simulations and their evolution with the true input energy.
\\
\textit{AGILESim} features a sligtly better angular resolution with respect to the \textit{GAMS} values but always consistent within a $2\sigma$ uncertainty level. We established in general a good agreement between the two simulators, especially if the \textit{FWHM} values are compared.


\subsection{In-flight PSF validation}
We achieve the scientific validation of the \textit{AGILESim} framework by comparing the simulated PSF profile with the angular resolution obtained from AGILE/GRID observations of the Vela pulsar.

\subsubsection{Vela data analysis}
\begin{figure*}
   \centering
   \includegraphics[width=0.53\textwidth]{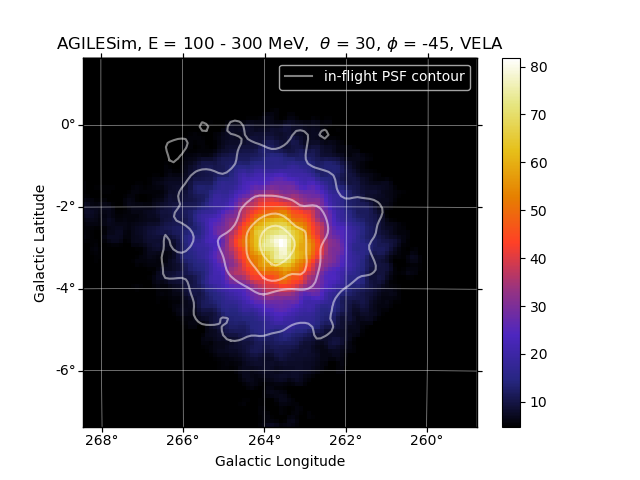}
   \includegraphics[width=0.46\textwidth]{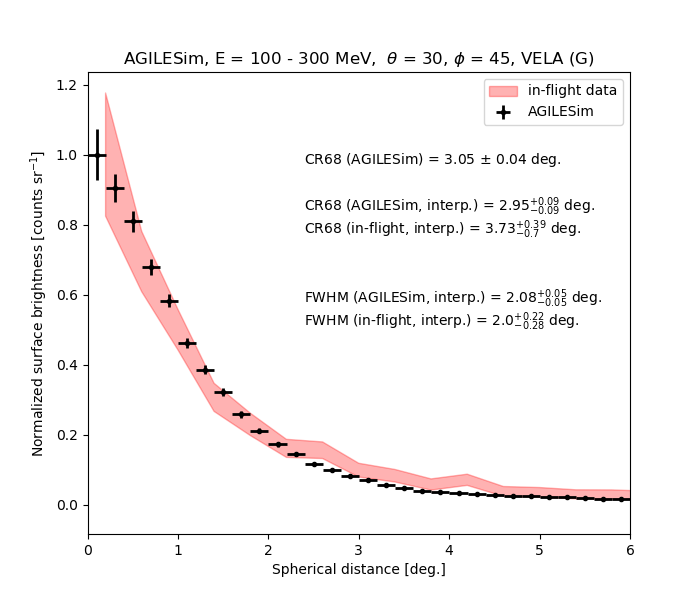}\\
 \includegraphics[width=0.53\textwidth]{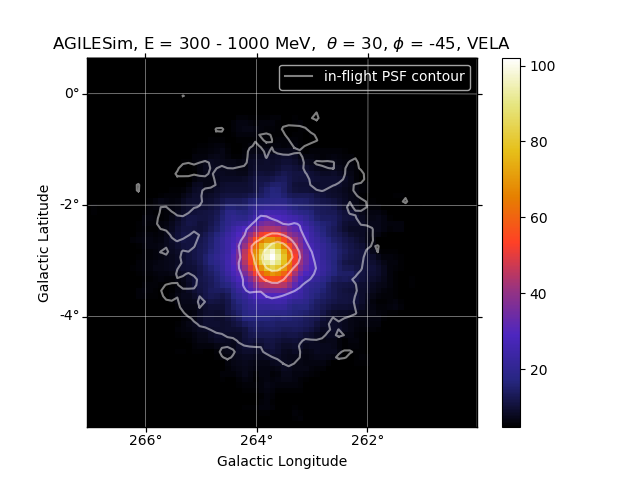}
   \includegraphics[width=0.46\textwidth]{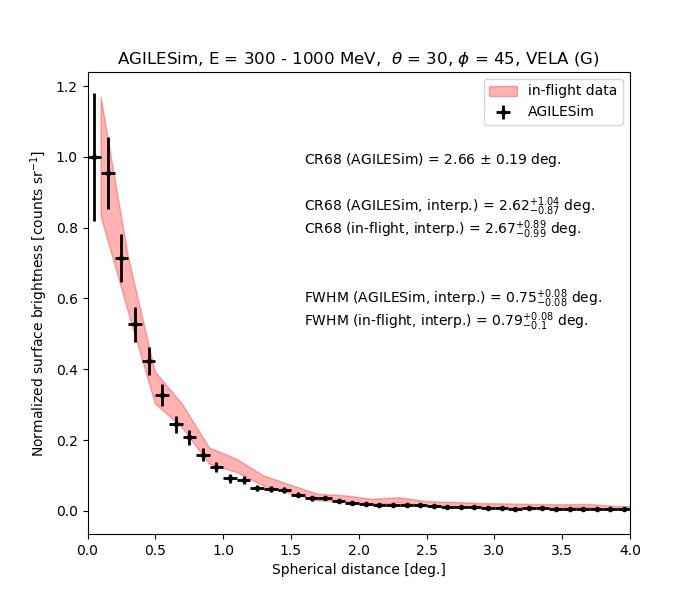}
   \caption{\label{fig:vela_psf_sim} \textit{Left panels}: Simulated counts map of the Vela pulsar, with in-flight PSF contours superimposed in white, in the 100 -- 300 MeV (top) and 300 -- 1000 MeV (bottom) energy ranges. \textit{Right panels}: Simulated (in black) and in-flight (in pink) PSF radial profile of the Vela pulsar in the 100 -- 300 MeV (top) and 300 -- 1000 MeV (bottom) energy ranges.}
 \end{figure*}
The Vela pulsar ($l = 263.59^{\circ}$, $b = -2.84^{\circ}$) is the brightest non-flaring gamma-ray source in the sky and one of the first targets observed by AGILE \citep{2009ApJ...691.1618P} and Fermi \citep{2010ApJ...713..154A}.
With no other sources near it and with a steady spectral distribution, the Vela pulsar is an optimal point-like target for tracking gamma-ray telescopes such as AGILE and Fermi and represents a good candidate for the characterization of the GRID angular resolution.
\\
From the second AGILE catalogue of gamma-ray sources \citep{2019A&A...627A..13B}, the Vela pulsar (2AGL J0835-4514) phase-averaged gamma-ray emission is best modeled by a power-law with a super exponential cut-off in the form:
\begin{equation}
\rm \frac{dN}{dE} = N_{0}E^{-\alpha}\;exp\left[-\left(\frac{E}{E_{\rm c}}\right)^{\beta}\right]\;,
\end{equation}
with $\alpha=1.71$, $\beta = 1.3$ and $E_{c}=3.91$ GeV.
The validation is performed in the core of GRID operative energy range (100 -- 1000 MeV), divided into two wide channels (100 -- 300 MeV and 300 -- 1000 MeV). 
\\
We extracted all observations of Vela with $\theta$ in the $20^{\circ}-40^{\circ}$ interval to be consistent with simulations obtained at an inclination $\theta=30^{\circ}$. The observation period covers almost two years, from December 1, 2007 (the end of the science verification phase) to October 15, 2009, the end of the AGILE operations in \textit{pointing} mode. The same data-set was used for the Vela analysis of the AGILE second catalogue. To further reduce gamma-ray albedo contamination from the Earth, the data selection excludes photons coming within $80^{\circ}$ from the reconstructed satellite-Earth vector.
\\
Counts from the Vela region are extracted in aperture photometry, by selecting from the \textit{G-flagged} event list all photons with reconstructed energy included in the intervals 100 -- 300 MeV and 300 -- 1000 MeV, and encircled in rings centered at the source, each with bin size of $0.1^{\circ}$. For each bin, a Poisson statistical error is assumed.
The surface brightness for each bin is obtained by dividing the counts in the ring for the subtended solid angle. Since Vela observations are not normalized in exposure, we do not take into account the effective area and the energy dispersion of AGILE/GRID to compare the observation with the output of the \textit{AGILESim} processing pipeline.
   \begin{table*}
\begin{center}
\begin{small}
\begin{tabular}{c|c|c|c|c|c}
\multicolumn{6}{c}{\textsc{PSF - Vela pulsar ($\theta=30^{\circ}$, $\phi=45^{\circ}$)}}\\
\hline
\hline
\multirow{4}{*}{} & \multicolumn{3}{c|}{\multirow{2}{*}{AGILESim}} & \multicolumn{2}{c}{\multirow{2}{*}{AGILE/GRID}}    \\
  & \multicolumn{3}{c|}{} & \multicolumn{2}{c}{}\\
\cline{2-6}
Energy & \multicolumn{1}{c|}{\multirow{2}{*}{analytical}} & \multicolumn{2}{c|}{\multirow{2}{*}{interpolation}}  & \multicolumn{2}{c}{\multirow{2}{*}{interpolation}}\\
& \multicolumn{1}{c|}{}& \multicolumn{2}{c|}{} & \multicolumn{2}{c}{}\\
\cline{2-6}
[MeV]  & \multirow{2}{*}{CR68}& \multirow{2}{*}{ CR68}& \multirow{2}{*}{FWHM}& \multirow{2}{*}{CR68} & \multirow{2}{*}{FWHM}\\
& & & & &  \\
& [$^{\circ}$] & [$^{\circ}$] & [$^{\circ}$]& [$^{\circ}$]&[$^{\circ}$] \\
\hline
\multirow{2}{*}{100 -- 300} &  \multirow{2}{*}{$3.1 \pm 0.04$}  & \multirow{2}{*}{$3.0_{-0.1}^{+0.1}$}&  \multirow{2}{*}{$2.1_{-0.1}^{+0.1}$} & \multirow{2}{*}{$3.7_{-0.7}^{+0.4}$}&   \multirow{2}{*}{$2.0_{-0.3}^{+0.2}$}\\
& & & & &  \\
\hline
\multirow{2}{*}{300 -- 1000} &  \multirow{2}{*}{$2.7 \pm 0.2$}  & \multirow{2}{*}{$2.6_{-0.9}^{+1.0}$}&  \multirow{2}{*}{$0.8_{-0.1}^{+0.1}$} & \multirow{2}{*}{$2.7_{-1.0}^{+0.9}$}&   \multirow{2}{*}{$0.8_{-0.1}^{+0.1}$}\\
& & & & &  \\
\hline
\end{tabular}
\end{small}
\end{center}
\caption{\label{tab:vela_psf}
Comparison of the AGILE/GRID PSF values obtained with the simulation (first three columns) and from the in-flight observation of the Vela pulsar (last two colums) in the 100 -- 300 MeV and 300 – 1000 MeV energy ranges.}
\end{table*}
\\
The in-flight count maps in the 100 -- 300 MeV and 300 -- 1000 MeV energy bands of the Vela pulsar are displayed in Fig. \ref{fig:vela_map_real}. A bin size of $0.1^{\circ}$ and a Gaussian smoothing are applied to both maps.  
The background surface brightness, in counts sr$^{-1}$, is extracted from an annular region at $8^{\circ}-10^{\circ}$ radial distance outside the AGILE/GRID PSF, and subtracted from the \textit{on-source} surface brightness. The resulting background subtracted radial profiles, with a binning of $0.4^{\circ}$ (100 -- 300 MeV) and $0.2^{\circ}$ (300 -- 1000 MeV) are shown in Fig. \ref{fig:vela_psf_real}. The error bars are statistical.
\\
A single King function well describes both profiles ($\chi^{2}/dof < 1.4$) although, because of the large statistical errors in the bins and the reduced number of bins, we obtain large errors in the best fit parameters. For this reason, the \textit{CR68} will be only evaluated by analytical and interpolation methods.

\subsubsection{Comparison with simulations}

Since the Vela pulsar exponential cut-off is well above the energy upper limit of our analysis, the source is simulated in \textit{AGILESim} as a simple power-law with spectral index $\alpha=1.71$ for a $\theta= 30^{\circ}$ inclination. We simulate the Vela pulsar with a wider energy range, from 30 MeV to 3 GeV, and select photon with a reconstructed energy in the 100 -- 300 and 300 - 1000 MeV channels, to take into account a potential contamination from photons with a true energy outside the energy range of analysis. The simulated count maps with superimposed the contours of the in-flight PSF for the two energy ranges are shown in the left panels of Fig. \ref{fig:vela_psf_sim}. The same binning and Gaussian smoothing of the in-flight maps are applied. The radial profiles of simulated and in-flight Vela pulsar observations are compared in the right panels of Fig. \ref{fig:vela_psf_sim}.
The \textit{CR68} and \textit{FWHM} are obtained by interpolation of the PSF profile, with the former also computed analytically from the list of background free photons provided by the simulation. All the values are listed in Table \ref{tab:vela_psf}.
\\
We find an in-flight angular resolution (\textit{FWHM}) for Vela-like sources of ${2.0^{\circ}}_{-0.3^{\circ}}^{+0.2^{\circ}}$ (100 - 300 MeV) and ${0.8^{\circ}}_{-0.1^{\circ}}^{+0.1^{\circ}}$ (300 - 1000 MeV). Our values are consistent with the ones reported in \citet{2015ApJ...809...60S} for Crab-like sources, $(2.5\pm0.5)^{\circ}$ (100 - 400 MeV) and $(1.2\pm0.5)^{\circ}$ (400 - 1000 MeV), where a higher spectral index $\alpha=2.1$ decreases the angular resolution.
\\
Since the \textit{CR68} takes into account the full distribution, the presence of large tails because of the wide energy channels causes large uncertainties in the \textit{CR68} obtained from interpolation, especially above 300 MeV where the statistic is lower. The \textit{FWHM} represents instead an unbiased measure of the angular resolution. We obtain a good agreement between the \textit{AGILESim} and the observed angular resolution of AGILE/GRID, in shape of radial profile and its measurables.

\section{\textit{BoGEMMS} and the future Gamma-ray telescopes}
At the dawn of the multi-messenger astronomy, after the discovery of the first electromagnetic counterpart of a gravitational wave event \citep{2017ApJ...848L..12A} and neutrino emission from a blazar \citep{2018Sci...361.1378I}, a continuous observational covering of the high energy sky has never been more urgent. 
Particle acceleration processes in relativistic jets of blazars, gas dynamics and nucleosynthesis studies by means of line diagnostic, transient phenomena (e.g. Gamma-ray Bursts) and new serendipitous sources are only some of the science objectives that could be unveiled by observations from hundreds of keV to tens of MeV.
\\
The e-ASTROGAM \citep{2017ExA....44...25D} and the NASA/AMEGO\footnote{https://asd.gsfc.nasa.gov/amego/index.html} mission concepts aim to increment of more than a factor 10 the COMPTEL sensitivity in the $0.2 - 30$ MeV energy range, de-facto opening a new observational window at the transition between the Compton and the pair production regimes. The two proposed gamma-ray telescopes are based on electron tracking technology for gamma-ray detection by means of Compton scattering and pair conversion interactions.
The main tracker design changes with respect to the AGILE/GRID and FERMI/LAT mass models are the removal of the high-Z converter, the minimization of the passive material in the trays and the use of double-sided Silicon strip detectors (DSSDs) in the planes. Such updates decrease the electron/positron multiple scattering, hence lowering the minimum detectable energy, while allowing the tracking of recoil electron produced by the Compton scattering. The lower attenuation efficiency of a single tray is compensated by the larger number of planes (56 for e-ASTROGAM). 
\\
A key ingredient in the sensitivity of such missions is the ability to discriminate between Compton and pair interactions and apply dedicated reconstruction algorithms accordingly. The correct simulation of both physics interaction, the charge collection, read-out and trigger logic is mandatory not only for the evaluation of the instrument performance but most importantly for their improvement.
The \textit{BoGEMMS} gamma-ray branch allows, with simple modifications of a configuration file, to build the mass model of any future electron tracking telescopes, including the Silicon tracker, the calorimeter and the anti-coincidence system, and to apply the post-processing and filtering criteria as a real observation in space. Testing its framework by reproducing on-ground and in-flight data scientifically validates its application to new missions sharing the same tracking concept.

\section{Summary}
The \textit{AGILESim} Monte Carlo simulator of the AGILE/GRID instrument, built with the \textit{BoGEMMS} Geant4-based framework, is first verified against pair conversion efficiency computed with tabulated data and on-ground measurements of the read-out strip charge distribution, then scientifically validated by successfully reproducing the AGILE/GRID angular resolution obtained from in-flight observations of the Vela pulsar in the 100 - 1000 GeV energy range. 
We obtain an in-flight PSF for Vela-like sources (power-law with spectral index $\alpha=1.71$) of ${2.0^{\circ}}_{-0.3^{\circ}}^{+0.2^{\circ}}$ (100 - 300 MeV) and ${0.8^{\circ}}_{-0.1^{\circ}}^{+0.1^{\circ}}$ (300 - 1000 MeV), in \textit{FWHM} of the radial profile. These values are in good agreement with the ones simulated by the \textit{AGILESim} pipeline, after verification with the mono-energetic PSF distribution obtained with the mission Monte Carlo simulator.
Along the validation process, we obtained that:
\begin{itemize}
\item the Geant4 models for Coulomb multiple scattering of leptons allow a very good representation of the charge sharing distribution;
\item the 68\% containment radius for real data, either form interpolation or from best-fit procedures, can be biased by fluctuations at the tail of the distribution;
\item if a fitting or interpolation procedure is not available, the \textit{FWHM} measure should be preferred because it is not influenced by the fluctuations in the tail;
\end{itemize}
The gamma-ray branch of the \textit{BoGEMMS} framework represents a reliable and user-friendly tool for the simulation of electron-tracking gamma-ray space telescopes.

\section*{Acknowledgements}
The AGILE Mission is funded by the Italian Space Agency (ASI) with scientific and programmatic participation by the Italian Institute of Astrophysics (INAF) and the Italian Institute of Nuclear Physics (INFN). Investigation supported by the ASI grant I/089/06/2. The authors wish to thank all the AGILE team for their help in producing this work.

\bibliographystyle{aasjournal}       
\bibliography{fioretti_agilesim} 



\end{document}